\documentclass[12pt,preprint]{aastex}

\newcommand{\cxo}{{\sl Chandra}}
\newcommand{\ngc}{NGC~5774/5775}

\newcommand{\xmm}{{\sl XMM-Newton}}
\newcommand{\hst}{{\sl Hubble}}
\newcommand{\msun}{M$_{\odot}$}
\newcommand{\ergl}{ergs~s$^{-1}$}

\newcommand{\ergcms}{ergs~cm$^{-2}$~s$^{-1}$}
\newcommand{\fxfo}{$F_{\rm X}/F_{\rm O}$}
\newcommand{\ha}{H$\alpha$}

\newcommand{\etal}{et al.}

\begin{document}

\title{ Multiwavelength study of the bright X-ray source population in the interacting galaxies NGC~5774/NGC~5775}

\author{
Kajal~K.~Ghosh\altaffilmark{1},
Douglas~A.~Swartz\altaffilmark{1}, 
Allyn~F.~Tennant\altaffilmark{2}, 
Lakshmi~Saripalli\altaffilmark{3},
Poshak Gandhi\altaffilmark{4},
C\'{e}dric Foellmi\altaffilmark{5},
Carlos M. Guti\'{e}rrez\altaffilmark{6} and
Martin L\'{o}pez-Corredoira\altaffilmark{6}
}
\altaffiltext{1}{Universities Space Research Association,
NASA Marshall Space Flight Center, VP62, Huntsville, AL, USA}
\altaffiltext{2}{Space Science Department,
NASA Marshall Space Flight Center, VP62, Huntsville, AL, USA}
\altaffiltext{3}{Raman Research Institute, Bangalore, India}
\altaffiltext{4}{RIKEN Institute of Physical \& Chemical Sciences, 2-1 Hirosawa, Wakoshi, Saitama 351-0198, Japan}
\altaffiltext{5}{Laboratoire d'Astrophysique, Universite Joseph-Fourier, 414 rue de la Piscine, 38400 Saint-Martin d'Heres, France}
\altaffiltext{6}{Instituto de Astrofísica de Canarias (IAC), E-38205 La Laguna, Tenerife, Spain}

\begin{abstract}
A few nearby interacting galaxies are known that host elevated number of ultraluminous X-ray sources. Here we report the results of a multiwavelength study  of the X-ray source population
 in the field of the interacting pair of 
 galaxies \ngc.
A total of 49 discrete sources are detected, including 12 
 ultraluminous X-ray source candidates with luminosities
 above $10^{39}$~\ergl\ in the 0.5~--~8.0~keV X-ray band.
X-ray source positions are mapped onto optical and radio images to search
 for potential counterparts. 
Twelve  sources in the field have  optical counterparts. 
Optical colors are used to differentiate these sources, which are mostly located outside the optical extent of the interacting  galaxies, as potential
 globular clusters (2), one compact blue dwarf galaxy  and   quasars (5). 
We obtained optical spectra of two of the latter, which confirm that they are background quasars. We expect 3 background sources in the field of these two galaxies.
These results are used to determine the true X-ray population of these two interacting galaxies, which are
connected with two bridges. Two high mass X-ray binaries are detected on these two bridges suggesting their formation through the interaction-induced  starformation episode.

NGC~5774 is an extremely low starforming galaxy with five X-ray sources plus three ultraluminous X-ray source candidates. Observed X-ray population of this galaxy does not scale  with the starformation rate alone but it may scale jointly with the mass of the galaxy and the starformation rate. 

Twenty four  X-ray sources (excluding the AGN) are detected in NGC~5775. and its X-ray luminosity function is consistent with that of other interacting galaxies, suggesting that these galaxies  have comparable  numbers of luminous sources. No X-ray point source was detected at the center of this galaxy to a limiting luminosity of 3$\times$10$^{37}$~ergs~s~$^{-1}$. Wind/outflow is detected from the central region of NGC~5775. Sub-solar diffuse gas with  temperature $\sim$0.31$\pm$0.04~keV is present in this galaxy, which suggest that NGC~5775
is in the beginning of the evolutionary process.

Twelve ultraluminous X-ray source candidates are detected within the D$_{25}$ isophotes of NGC5774/5775.
Several of them  are highly variable X-ray sources that fall below the
 detection levels in one of two X-ray observations spaced 15 months apart. Two ultraluminous X-ray sources are located in the halo of NGC5775 and one of theme is hosted in a globular star cluster. Four of the remaining 10 candidates have powerlaw X-ray spectra with photon indices around 1.8 and are extremely luminous with no optical counterparts.  One of these four objects  is the brightest ($\sim$10$^{41}$~\ergl) with a possible 6.2 hr period  and it varied  by more than a factor of 500. 
Two of the rest six ultraluminous X-ray source candidates are having steep-powerlaw X-ray spectra and are embedded in diffuse H$\alpha$ emission, which are probably ionized nebulae. These nebulae could be due to energetic supernova explosions or to continuous inflation by jets. Rest four ultraluminous X-ray source candidates are flat-powerlaw X-ray sources hosted in either young star clusters or bright starforming complexes. Two of them are radio sources. Finally, we find that the number of ultraluminous X-ray source candidates in interacting/merging galaxies are correlated with the FIR, K-band and UV luminosities of their host galaxies, suggesting that the formation and evolution of ultraluminous X-ray sources  depend not only on the starformation rate but also on the mass of their host galaxies.
\end{abstract}


\section{Introduction}

The most luminous non-nuclear X-ray sources in nearby galaxies
 occur in regions of high rates of current star formation (Roberts  \& Warwick  2000; Colbert \& Ptak 2002; Swartz et al 2004; Liu \& Mirabel 2005). 
Whether these ultra-luminous X-ray sources (ULXs) are 
 powered by some exotic emission process 
or represent the extreme end of stellar-mass 
 black hole formation  
 (in excess of $\sim$20~\msun,  the upper limit for stars
 of moderate metallicity evolving in isolation, Fryer \& Kalogera 2001) 
 is a matter of current debate (for recent reviews on ULXs see Fabbiano 2006; King 2006; Fabbiano \& White 2006).
Different models have been proposed to explain  the ULX phenomena and they can be grouped into different broad categories: (i) geometrically/mechanically- (King et al. 2001a; Fabrika \& Mescheryakov 2001; Fabrika 2004; Poutanen et al. 2007) or relativistically beamed or super-Eddington accretion stellar-mass black hole systems (K\"{o}rding \etal\ 2002; Georganopoulus \etal\ 2002; Abramowicz et al. 1980; Arons 1992; Gammie 1998; Begelman 2002, 2006; Grim, Gilfanov \& Sunyaev 2002), (ii) supernovae and hypernovae (Terlevich 1992; Schlegel 1995; Paczynski 1998; Wang 1999; Li 2003), (iii) supersoft sources (Swartz  et al. 2002; Kong \& Di Stefano 2005),
(iv) accreting intermediate-mass black hole systems (IMBHs) with masses in the range 50~--~10$^{5}$~M$_{\odot}$ (Colbert \& Mushotzky 1999; Makishima  et al. 2000; Madau  \& Rees  2001; Ebisuzaki  et al. 2001; Portegies-Zwart   \& McMillian  2002; Miller  \& Hamilton  2002; Ho, Terashima \& Okajima 2003; Mushotzky 2004; van der Marel 2004; Freitag et al. 2006),  and (v) foreground/background objects, which mimic as ULXs (Arp et al. 2004;  Gutierrez 2006; Gutierrez \& Lopez-Corredoira  2005, 2007 and references therein).

Observationally, periodic variations in the X-ray light curves of some ULXs have been detected, which suggest that they are stellar-mass black hole binaries (Sugiho et al. 2001; Bauer et al. 2001; Liu et al. 2002, 2005; Strohmayer \& Mushotzky 2003; Pietsch, Haberl \& Vogler 2003; Pietsch et al. 2004; Stobbart, Roberts \& Warwick 2004; Weisskopf et al. 2004; Soria et al. 2004; Soria \& Motch 2004; Mukai et al. 2005; Ghosh et al. 2006; Fabbiano et al. 2006). In addition, spectral 
curvatures have  been detected in the X-ray spectra of a few luminous ULXs (Dewangan et al.  2004; 2006 and references therein), which suggest that these objects are beamed stellar-mass black hole binaries. On the other hand, 
by analogy with the X-ray spectra of Galactic black hole binaries, a 
multicolor disk (MCD) blackbody component with 
temperature around 100 eV (plus powerlaw) has been considered as the signature of a cool accretion disk of an 
IMBH system (Miller et al. 2003; Kaaret et al. 2003; Miller et al. 2004; 2004a; Cropper et al. 2004; Kong et al. 2004). In addition,
quasi-periodic oscillations (QPOs) with long quasi periods have been detected 
in some ULXs (Liu et al 2005; Strohmayor \& Mushotzky et al. 2003; Soria et al. 2004; Strohmayer et al. 2007). Comparison of 
these results with the scaling relation between black hole mass and break-frequency of QPO  suggests the
presence of IMBHs in these ULXs (Belloni \& Hasinger 1990). 
However, the  suggestions of IMBH systems based on the MCD model have been questioned and alternate explanations in the frame works of stellar-mass black 
hole systems have been proposed (King \& Pounds 2003;  Roberts et al. 2005; Stobbart et al. 2006; Goncalves \& Soria 2006; Barnard et al. 2007). 
In addition, Hui, Krolik \& Hubeney (2005) have shown that non-LTE accretion flows around IMBHs can easily generate hot accretion disks with temperatures up to 1~keV, which is in sharp contrast to the conventional MCD model. 
All these results suggest that ULXs are heterogeneous objects  (Feng \& Kaaret 2005, 2006; Roberts et al. 2006; Winter et al. 2006, 2007).
Similarly, optical photometric and spectroscopic studies have revealed that stars, star clusters, ionized nebulae, etc. are possible optical counterparts of ULXs (Ghosh et al. 2001, 2005; Soria et al. 2005; Ptak et al. 2006; Ramsey et al. 2006). Optical spectroscopic studies on the local environments of some ULXs have detected  different emission lines, which suggest that both shock  and photo-ionization excitations  are the dominant physical processes around these objects (Abolmasov 2007a,b; Mucciarelli et al. 2005, 2007; Zepf et al. 2007). {\it In summary, no clear picture is emerging to differentiate between compact accretor models of ULXs, yet it is certain that most are young objects born in extreme environments}.

Some of the highest star formation rates are induced by
 interactions among gas-rich (spiral) galaxies and it is natural to
 expect these interacting galaxies to host numerous ULXs.
This is consistent with the observed anti-correlation between the 
 number of ULXs per galaxy and nearest neighbor distance (Swartz \etal\ 2004).
The face-on
 pair of merging galaxies NGC~4038/4039 (the Antennae)
 is the prototypical example of this paradigm: 
As many as 18 ULXs are present in the Antennae
 depending on the distance adopted
 and defining ULXs to have X-ray luminosities above $10^{39}$~\ergl\
 (Zezas et al. 2002). In addition, we have found from our survey of ULXs in the \cxo\ archive 
of galaxies that there is a correlation between the number of ULXs and the far infrared luminosity
of interacting galaxies (Swartz et al. 2004). Presently, there are a few known interacting galaxies
that host elevated number of ULXs. Thus, X-ray studies of interacting galaxies are important, specially from
the perspective of ULX studies.

The galaxy pair \ngc\ is another example of star formation 
 triggered by tidal interaction between galaxies
 resulting in the creation of many luminous X-ray sources. 
NGC~5775 is a SBc starburst galaxy seen nearly edge-on, $i\sim86\arcdeg$,
 with star formation extending throughout its visible disk.
As with many interacting pairs (e.g., Kennicutt \etal\ 1987), 
 star formation in NGC~5775,  although triggered by external forces,
 appears to result from tidally
 induced gas motions within the disk of NGC~5775 rather than by 
 mass transfer between the galaxies.
It is connected to NGC~5774 by H{\sc i} bridges (Irwin 1994) but
 there are no other obvious distortions in optical images such as 
 tidal tails or disk warping common in more advanced stages of interaction
(for example, tidal tails are the defining optical characteristics of the 
 Antennae merging galaxy pair).
The outer disk of the face-on companion SAB(rs)d galaxy NGC~5774 may have been
 disrupted by the interaction (Irwin 1994) and this 
 is the apparent source of the bridge material. 
The rate of mass transfer through the bridge is estimated
 to be only about 5--10 \msun\ yr$^{-1}$ and
Irwin (1994) estimates the age of the interaction from 
 dynamical considerations to be $\sim$10$^8$~yr. 
Star formation in NGC~5774 is proceeding at a normal rate for an Sd galaxy.
Neither galaxy displays strong AGN activity (Ho, Filippenko \&
 Sargent 1997).
 
The \cxo\ X-ray Observatory image of the \ngc\ pair reflects the
 contrasts between the IR-bright starburst galaxy NGC~5775 and its relatively
 normal companion NGC~5774. Only nine sources are detected in NGC~5774 and there is 
 no diffuse component visible.
In contrast, twenty-five bright discrete sources are visible above the detection limit
 of $\sim$10$^{38}$~\ergl\ in the disk of NGC~5775.
The disk sources are embedded in extensive diffuse hot gas.
Several more sources are in the surrounding region including
 the H{\sc i} bridges. 
Thus, this interacting pair of galaxies, \ngc\, is an ideal laboratory to investigate the formation of
X-ray sources on the bridges, to study the role of starburst activities on the evolution of galaxies
and formation of ULXs, and finally to study the nature of a large number of wide varieties of ULXs.
To achieve these scientific objectives,
we have analyzed the available X-ray, optical, and radio data of the
 galaxy pair and their surroundings; supplemented by new
 optical imaging, 
 spectroscopy, and radio continuum 
 measurements. In \S~2, we describe the observations, our data reduction methods and
results. In \S~3, 
 we discuss the source population as a whole in the context of current
 theories for ULX formation and the relationship between ULXs and galactic
 dynamics and star formation. Conclusions of the present studies are listed in \S~4.
We adopt a cosmology  H$_{0}$=73~km~s$^{-1}$~Mpc$^{-1}$, $\Omega_{M}$=0.24 and $\Omega_{\Lambda}$ =0.76
(Spergel et al. 2007).

\section{Multiwavelength Observations and Data Analysis}

An optical image of the \ngc\ field is shown in Figure~1. 
The image is 6.\arcmin 6$\times$5.\arcmin 8 corresponding to 
 51.5$\times$45.2 kpc at the 26.8~Mpc distance to the galaxy pair (Tully 1988).
The H{\sc i} bridges connecting these two galaxies extend roughly along
 the line described by the disk of the edge-on starburst galaxy 
NGC~5775 from the NW tip of NGC~5775 
 to NGC~5774 and from the SE of NGC~5774 eastward to the center of NGC~5775.
The NW bridge is visible in the optical image, shown in Fig. 1 but the SE bridge is barely visible. However, they are much more pronounced
 at radio wavelengths as shown by Irwin (1994) and by Lee \etal\ (2001).

\subsection{X-ray Observations }

Chandra Advanced CCD Imaging Spectrometer (ACIS) observations of 
 the \ngc\ system were carried out on April 05, 2002 for 58.2 ks (ObsID 2940 \&  
  Principal Investigator -- Philip Maloney).
Level 2 event list for this data set was retrieved from the Chandra archive. 
NGC~5774 is located on the front-illuminated CCD S2 and NGC~5775 is
 on the back-illuminated CCD S3.
We use the locally-developed software tool LEXTRCT (Tennant 2006) for data analysis, which is described in details in Swartz et al. (2003).
Source detections in LEXTRCT were performed using a circular Gaussian approximation
to the point spread function (PSF), which gives higher
weight to sources with a central concentration of events. Point-source counts and spectra were extracted from
within the 95\% encircled-energy aperture of the model PSF.
Background regions were typically chosen from annular
regions surrounding the source regions except in crowded
regions of the field where background regions adjacent to
the source were used. The background-subtracted counts
within the source regions were scaled to obtain the aperture-corrected count values. The background-subtracted point
source detection limit is 14 counts for the 2.8 minimum signal-to-noise ratio (S/N)
threshold and a minimum 5 $\sigma$ above background.  We define the total field as the 6.\arcmin 6$\times$5.\arcmin 8
rectangle, in R.A. and Dec, that inscribes
 both  D$_{25}$ ellipses. Forty-nine sources were detected in this field.
 Their properties are listed in Table~1.
The \cxo\ image of the \ngc\ field is shown in Figure~2
with the distribution of X-ray sources detected in the field. 
The ellipses shown delineate the D$_{25}$ isophotes of these two galaxies.

Source and background spectra and light curves
were used for spectral and timing analysis, respectively.
X-ray light curves were binned into 1000~s bins. 
We  computed the Kolmogorov-Smirnov statistic to determine the time variability of sources by comparing the cumulative event arrival times, binned
 at the 3.24~s frame time, to that expected for a steady source.
The resulting KS probabilities are listed in column 10 of Table~1.  
We also performed  a $\chi^{2}$ test against a constant flux hypothesis for each
 light curve  and the results are presented in column 11 of Table~1.

X-ray spectra of all the sources and the corresponding backgrounds were 
 binned so that there were at least 10 counts per fitting bin. 
Spectral fit were made to sources with more than 40 detected source counts.
Spectral redistribution matrices and ancillary response 
 files were generated using \cxo\ X-ray Center software CIAO version 3.4. 
XSPEC version 11.3.2t was used to fit the 0.5 -- 8.0 keV energy spectra with 
 an absorbed ({\tt phabs}) powerlaw ({\tt powerlaw}) model. 
Best-fitting model parameters and fit statistics are listed in columns 5--7 of Table~1. 
These spectra have also been fitted with other models 
 as described in the sections on individual sources (\S 3 and \S 4).
The observed spectra, with best-fitting powerlaw models and fit residuals, 
 are shown in Figure~3. 
Intrinsic luminosities of these sources based on these spectral fits
 are listed in column 8 of Table~1. 
Intrinsic luminosities for sources with fewer than 40 source counts 
 were estimated by allowing only the model normalization to vary
 in the fitting procedure (using appropriate response matrices 
 and ancillary response files)
 and freezing the powerlaw photon index to 1.8 and the absorption column
 to the Galactic $N_{\rm H}$ value
 in the direction of \ngc\ (3.47 x 10$^{20}$~cm$^{-2}$) in WebPIMMS\footnote{http://heasarc.gsfc.nasa.gov/Tools/w3pimms.html}.
The photon index value of 1.8 is very close to the average value of the powerlaw index, for sources with statistically-acceptable powerlaw fits measured in our catalog of ULXs (Swartz et al. 2004).

  
\begin{center}
\includegraphics[angle=0,width=\columnwidth]{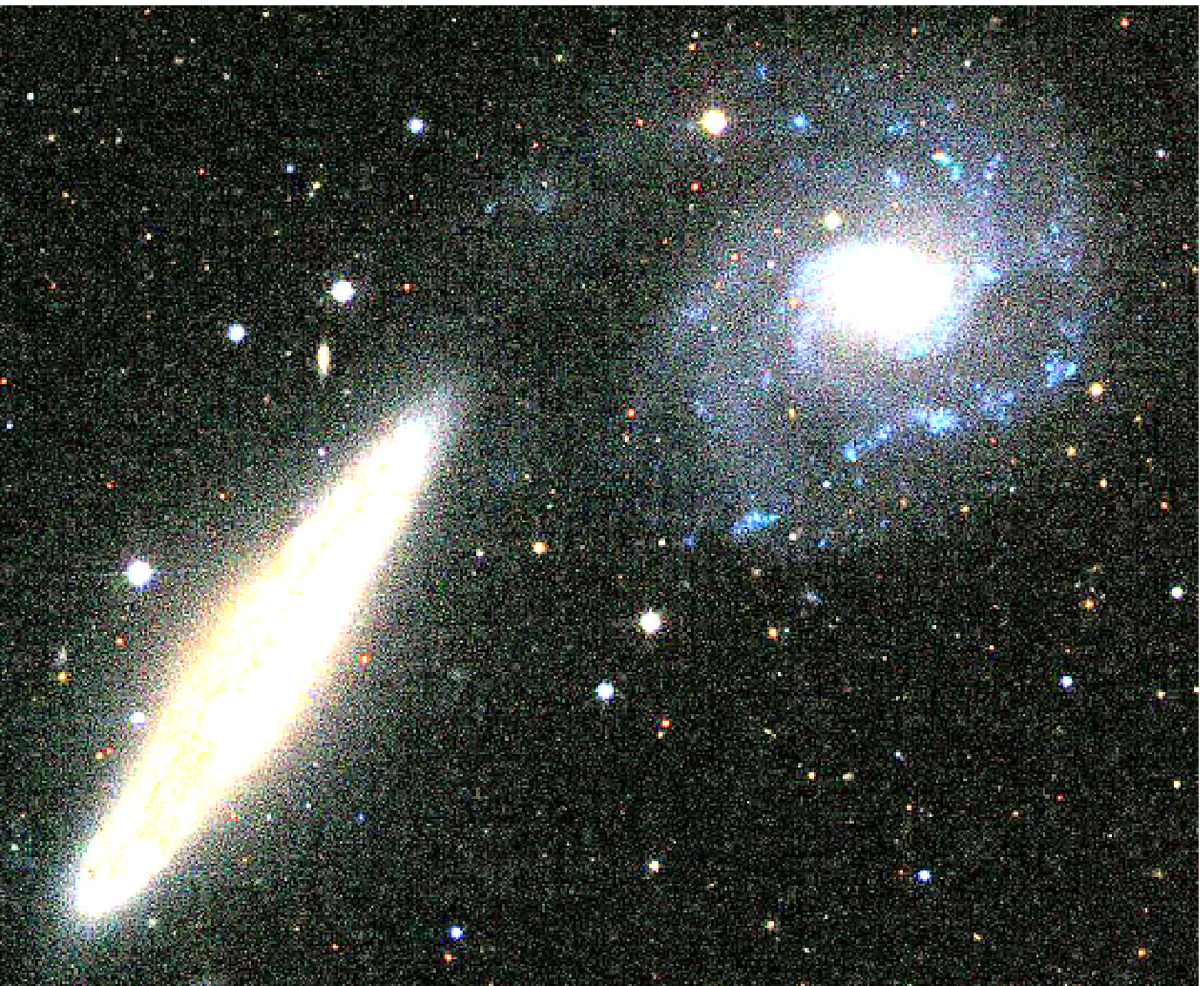}
\figcaption{
Optical image of the \ngc\ field adapted from Sloan Digital Sky Survey
data (blue, green and red correspond to the g, r and i bands). 
North is up and East is to the left in this and all subsequent images of the 
 \ngc\ field.
The starburst galaxy NGC~5775 is the nearly edge-on galaxy to the left (east) in this image, NGC~5774 is nearly face-on and located to the NW.  Diffuse optical emission is present between NGC~5774 and NGC~5775, 
visible to the north-east of NGC~5775, 
where an H{\sc i} bridge connects these two galaxies. Another, less prominent,
bridge extends from the SE of NGC~5774 eastward to NGC~5775
(cf. Irwin 1994 Figure~3; Lee \etal\ 2001 Figure~1.), which will be defined as the SE-bridge.}
\end{center}



\begin{center}
\includegraphics[angle=-90,width=\columnwidth]{Fig2.ps}
\figcaption{\cxo\ 6.\arcmin 6$\times$5.\arcmin 8 X-ray image of \ngc\ field.
Positions of detected sources are marked with crosses and circles whose 
sizes are proportional to the size of the point-spread function at their
position on the detector. Twelve sources with possible optical counterparts are marked with squares.  Possible optical counterpart
means the presence of an optical object within the astrometric-corrected error circle at the position of the Chandra source.
Source numbering follows column 1 of Table 1.  
Ellipses denote the D$_{25}$ isophotes of the two galaxies. Note the excess diffuse emission in the disk of NGC~5775. North is up and East is to the left. Axes denote ACIS pixels ($\sim$0.\arcsec492 per pixel). Sources are numbered in order of increasing R.A. }
\end{center}

\begin{center}
\includegraphics[angle=0,height=22.0cm,width=18.0cm]{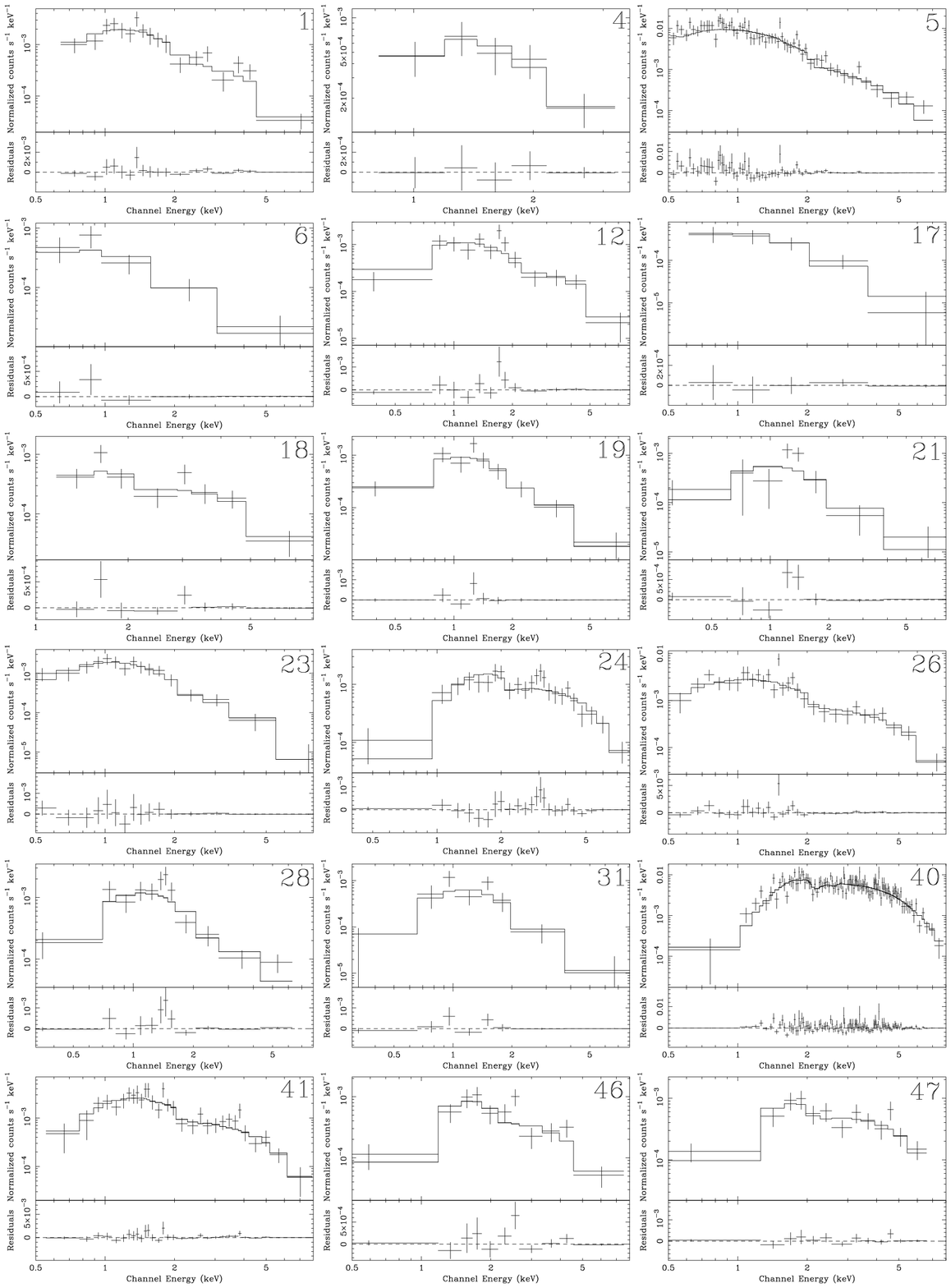}
\figcaption{The X-ray spectra of the 18 Chandra sources in the field of \ngc\ with more than 40 source counts detected (top panel). The curve through the data is the best-fitting powerlaw model. Fit residuals are shown in the lower panel for each source.}
\end{center}

A 47~ks \xmm\ observation of \ngc\ was carried out on July 2003 (ObsID~0150350101).
The data was processed using the \xmm\ Science Analysis System (version 6.5.0).
This observation suffered from background flares. After filtering these flares, only 22.6~ks of useful data remained.
The \xmm/MOS~1 
image of NGC~5774/5775 is shown in Figure~4. 
Seven sources were detected in the entire field.
Their X-ray properties, using the EPIC MOS and PN data, are listed in Table~2. 
Six of these were previously detected with \cxo\ 
 (and they are numbered in Table~2 using the numbering scheme of Table~1). 
The one source detected only in the XMM-Newton data is listed source~\#49 in Table~1. 
We did not perform variability studies on the \xmm\ data due
 to the strong background flare activity.  
Absorbed powerlaw model fit parameters to the spectra of these sources
 are listed in Table~2. Sources were detected down to a detection limit of 
$\sim$7~$\times$ 10$^{38}$ ~ergs~s$^{-1}$ for 14 counts in a source above the background in the 0.2--10.0 keV band, for the EPIC MOS and 2.0~$\times$ 10$^{38}$ ~ergs~s$^{-1}$ in PN observations, respectively.


\begin{center}
\includegraphics[angle=-90,width=\columnwidth]{Fig4.ps}
\figcaption{The \xmm/MOS~1 image of NGC~5774/5775 with the positions of detected sources marked with circles of 10\arcsec.0 radius. }
\end{center}


\begin{center}
\includegraphics[angle=0,height=22.0cm,width=18.0cm]{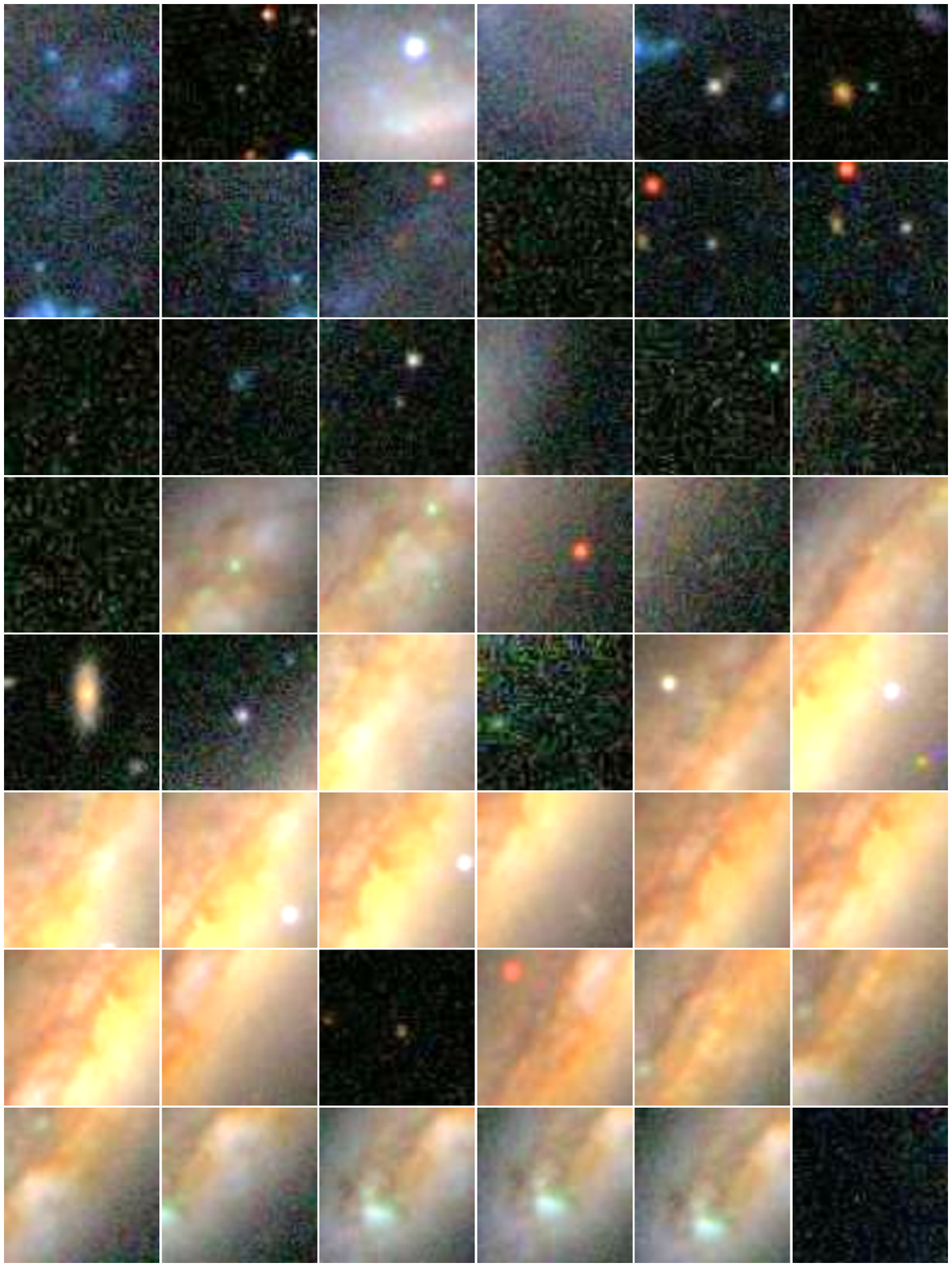}
\figcaption{25\arcsec $\times$25\arcsec\  SDSS 3-color images
centered on the \cxo\ sources in the field of \ngc. Colors are the same as used in Figure 1. North is up and east to the left.}
\end{center}

For completeness, we analyzed archival ROSAT observations of \ngc. 
A 6~ks PSPC observation was taken in July/August 1993 and
 a 35~ks HRI observation was taken in July 1997. 
No sources were detected in the PSPC data and only two sources 
 (\#5 and \#26 of Table~1) were detected in the HRI data. 
Count rates of these two sources were used to estimate their intrinsic fluxes  
 and are listed in Table~2.

\subsection{Optical Photometric Observations} \label{s:o_phot}

Photometric data of \ngc\ observed with the \hst\ Space Telescope, 
 the European Southern Observatory's 3.5~m 
 New Technology Telescope (NTT), and the
Sloan Digital Sky Survey (SDSS) program were obtained from their respective archives. 

A 640~s
\hst/WFPC2 observation of NGC~5774 was carried out on May 20, 2001 
 using the F814W filter. 
This observation covered  only X-ray source s01 in Fig.2. 
\hst\ astrometry was done using the USNO-B1.0 star 0935-0243054.
Based on the absolute positional uncertainties of the USNO stars and the 
X-ray sources, we estimate that the astrometric accuracy  was better than 
0.\arcsec70. No counterpart to source~\#1 was detected
in the \hst\ data. 

Two \hst/ACS--WFC observations of NGC~5775 were carried out on August 21, 2005 with F625W and F658N filters. Astrometry between \hst/ACS--WFC and Chandra  images were carried out using the source~\#26. Considering the centroiding  uncertainties
of these images, the estimated astrometric accuracies were in the range of 0.\arcsec2 -- 0.\arcsec3.

\ngc\ was observed with the  ESO/NTT
in \ha\ and R--band filters on May 07, 1991. 
Multiple short observations were taken in each filter.
The total integration time for the \ha\  filter (full-width at half-maximum =75 \AA) 
was 60 m, which reached to the {\it rms} noise level of 7$\times$10$^{-18}$ ~ergs~cm$^{-2}$~s$^{-1}$~arcsec$^{-2}$ (Lee et al. 2001).
Photometric analysis of these images were carried out using 
IRAF\footnote{IRAF is the Image Reduction and Analysis Facility,
written and supported by the IRAF programming group at the National Optical
Astronomy Observatories (NOAO) in Tucson, Arizona.} software.
Astrometry between the ESO/NTT and \cxo\ images was done using two sources (\#5 and 26), which are bright in both the images.

SDSS photometric ($u$, $g$, $r$, $i$, and $z$ bands) images of the field and spectroscopic 
observations of the nuclei of NGC~5774 and NGC~5775 
 were carried out on May 16, 2001. 
Astrometry was performed following the same procedure used for the ESO/NTT images. Astrometrically-corrected SDSS images 
(25\arcsec $\times$25\arcsec ), centered on the positions of 
the \cxo\ sources, are shown in Figure~5. The X-ray (0.5--8.0 keV) to optical (V band) flux ratios (F$_{X}$/F$_{O}$) are given in column 9 of Table~1. When no candidate
optical counterpart to an X-ray source is  detected, then the value of the V band has been assumed to be 25 mag and the corresponding lower limit values of F$_{X}$/F$_{O}$ are listed in Table~1. SDSS detection limit in the V-band is 25 mag.  For sources in confused region of NGC~5774/5775, no F$_{X}$/F$_{O}$ is given.

There are 12 
X-ray sources with potential counterparts in the SDSS images.
Their properties are listed in Table~3. Where objects are identified in the SDSS catalog, these magnitudes are quoted. For sources, \#9 and \#26, magnitudes were estimated by fitting normalized Gaussian profiles to the spatial distribution of counts in each filter image.
Most of these objects lie outside the D$_{25}$ isophotes of the galaxy pair.
Three of the X-ray sources, \#5, 23, and 26, are ULX candidates based on 
their X-ray flux and assuming they are at the distance of \ngc.
An optical color-color diagram including these 12 sources 
 is displayed in Figure~6. Usually, the nominal photometric accuracy of SDSS images is around 0.03 mag for sources not limited by photon statistics (Ivezi{\'c} et al. 2004). However, some of the 12 sources are relatively faint and we estimated that the photometric accuracies in $(u-g)$ and $(g-r)$  colors will be approximately between 0.10 and 1.0 mag\footnote{http://www.sdss.org/dr4/start/aboutdr4.html}.
Five sources (sources~\#5, 6, 9, 15, and 26) are located in a small region 
near the center  in Figure~6. As shown in \S~\ref{s:o_spectra}, optical 
 spectra of sources~\#5 and 26 show they are  background QSOs. 
The $(u-g)$ and $(g-r)$ colors of these sources are 
consistent with being $z<3.0$ SDSS-QSOs (Richards et al. 2001). We conclude all five
 sources 
are likely background QSOs with redshifts less than 3 but only have strong spectroscopic evidence for
this classification for sources \#5 and 26.


\begin{center}
\includegraphics[angle=-90,width=\columnwidth]{Fig6.ps}
\figcaption{(u-g) versus (g-r) plot of optical sources detected in the SDSS data at the astrometric corrected positions of Chandra sources. Long dashed line with arrows shows the positions of the source \#25 and the associated galaxy at a redshift of 0.0953 (see text).}
\end{center}

Three objects located in the lower right hand region of Figure~6
 could be globular clusters (GCs) based on their colors and absolute magnitudes (assuming they are at the distance of \ngc)
compared to those of Galactic and extragalactic GCs (e.g., Harris 1996; Miller \etal\ 1997; Lotz \etal\ 2004). Source \#23 is in the halo of NGC~5775 and source \#11 in the halo of NGC~5774 so this identification is
plausible. Source \#11 is located close to a background galaxy on the (u-g) versus (g-r) plot. However, source \#11 is a point source. Thus, we believe that it is a GC associated with NGC~5774. Source \#2 is about 2 D$_{25}$ radii from NGC~5774 and it may
instead be an unrelated background source.

Source  \#25 and 39 are peculiar in that their positions in the 
upper left of Figure~6 are due to  weak emission
in the $g$ band relative to the $u$ and $r$ bands. 

We reanalyzed the Sloan images of source \#39 and
find that the $u$ band emission is significant only at about the 1 sigma level.
Therefore we feel that the $u$ band flux listed in the Sloan catalog
is overestimated and that the tabulated colors are unreliable for 
this source.

Source \#25 is located at the southern tip of what is clearly an
elongated background galaxy, SDSS J145355.82+033431.8 (Figure~5).
Our interpretation of the Sloan image is that the X-ray source
is coincident with a  blue color  region different from the rest of
this galaxy. The SDSS color of the astrometric corrected X-ray position
 is located at the upper left  in Figure~6.
We were unable to improve on these  color estimates but feel that they
 are not reliable because of the spatial overlap.
We note that the colors of the background galaxy center (labeled ``Galaxy" in Figure 6) alone is
very close to the colors of source \#11, which is reasonable 
in that the colors of the old stellar population of a galactic bulge 
at low redshift should match the colors of a globular cluster.

The fact that source \#25 is not at the center of the background galaxy
suggests that it could be a very luminous ULX in that galaxy.
The Sloan spectrum of this galaxy gives a redshift of 0.0953$\pm$0.0002, which corresponds to
a cosmology corrected luminosity distance of 424 Mpc. The Galactic absorption corrected
flux for the source \#25 in the 0.2--10 keV band is 3.1$^{+2.0}_{-1.1}\times10^{-15}$~\ergcms
and the corresponding  luminosity  is $\sim$6.7$\times$10$^{40}$~\ergl. 
The immediate environment of this source in the host galaxy is blue, which can be seen from Fig. 6.
No HST data is available for this galaxy.
To search for the optical counterpart of source \#25, we added all the available R-band images obtained from ESO/NTT. 
Astrometric and photometric studies were performed with the final image. No counterpart within the error circle at the astrometric corrected
Chandra position of source \#25 was detected in this image, at the limit of 27.5~mag. This suggests that the X-ray to optical flux ratio (\fxfo) is more than 17. Absence of radio detections and such high value of \fxfo\ indicates that the source \#25 is not a background AGN and most likely a bright ULX in SDSS J145355.82+033431.8.

The  source~\#13 is weak and consequently has poorly-measured
 optical colors.
Formally, this source has colors similar to a star forming region
 but it is located far from any active star forming region so this 
 interpretation is doubtful.  

The remaining source~\#17 has not been cataloged in the SDSS. Our measurements are listed in Table~3 and shows that it is an extended object with extremely blue colors. This is a blue compact dwarf galaxy.

In summary, of 12 optical sources spatially coincident with X-ray sources 
two could be globular clusters (s11 and s23), one bright ULX (s25) in a background galaxy,  one blue compact dwarf galaxy (s17), six are likely background QSOs (s02, s05, s06, s09, s15 and s26), and two are too faint or too confused to identify with confidence (s13 and s39).

\subsection{Optical Spectroscopic Observations} \label{s:o_spectra}

Comparing Tables~1 and 3, we find three of the 15 
ULX candidates (source \#5, 23 and 26) have potential optical counterparts. Sources \#5  and 26 are brighter than 21 mag and we  obtained optical spectra of these two objects. 

\begin{center}
\includegraphics[angle=90,width=\columnwidth]{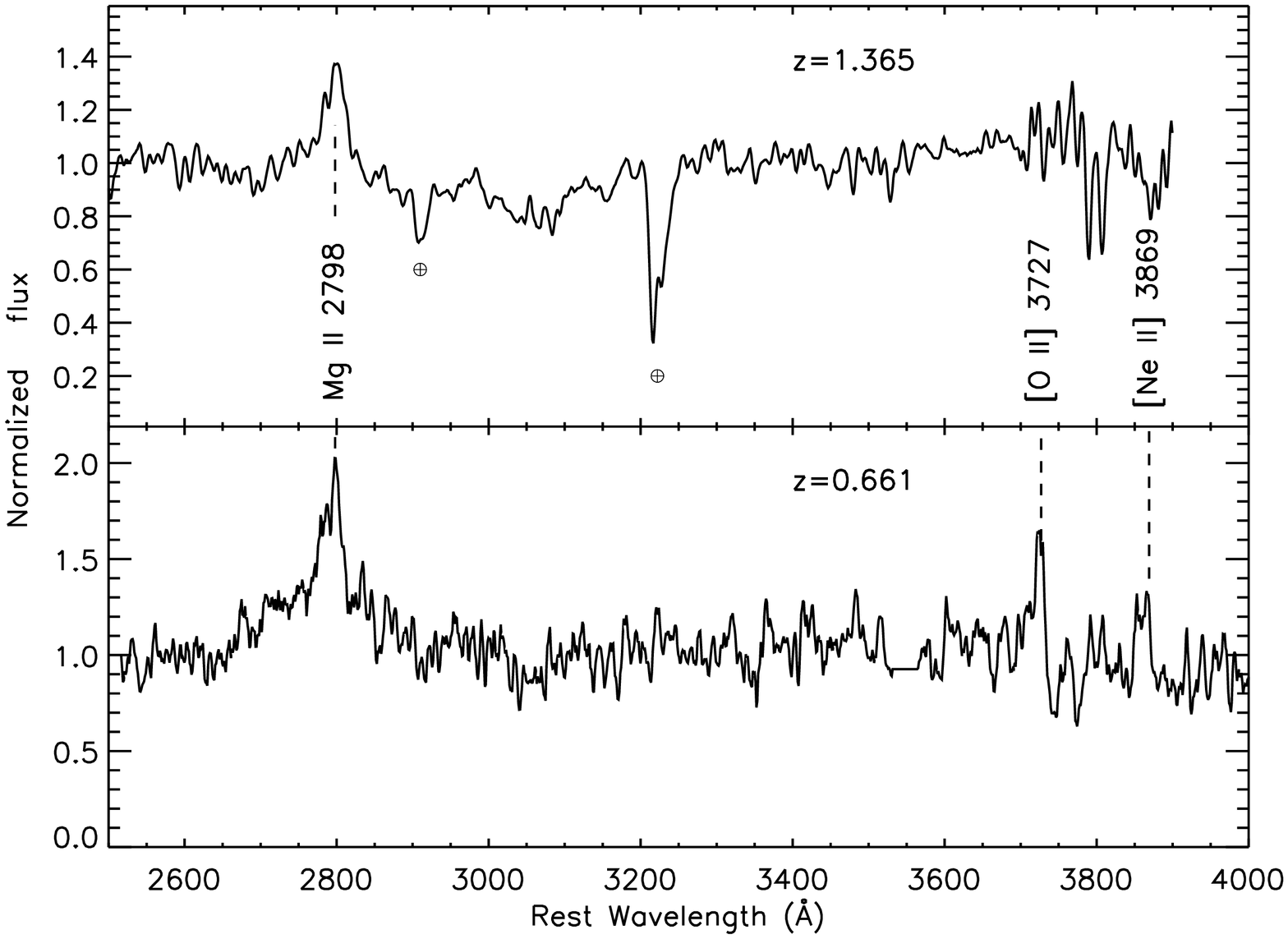}
\figcaption{Optical spectrum of source\#5 (at the source frame)  obtained  
with the WHT using the ISIS spectrograph. A strong emission line due to 
MgII $\lambda$2798~\AA\ from a background quasar at a redshift of 1.365 is
shown. The strong absorption line is due to telluric absorption in the Earth's atmosphere (top panel). Bottom panel shows the spectrum of source\#26 obtained with the  DOLORES spectrograph at
the 3.6~m Telescopio Nazionale Galileo telescope. This is the spectrum of a background quasar at a redshift of 0.661.}
\end{center}


The
optical spectrum of source~\#5 was obtained on January 2004 with the William Herschel Telescope (WHT)\footnote{The William Herschel Telescope  is operated by the Isaac 
Newton Group and the IAC in Spain's Roque de los Muchachos Observatory. The  observations were done in service time.}
using the red arm of the ISIS spectrograph and the
R158R grism during an 1800~s exposure. 
The slit width was between 1.\arcsec 20 and 1.\arcsec 45. 
Cu--Ar and Cu--Ne lamps were used for wavelength calibration. 
This  provides a sampling of 
1.62 \AA\ pixel$^{-1}$ and an effective resolution of 8--10 \AA\ 
(depending on the slit width used and on atmospheric seeing). 
The spectrum was analyzed following a standard
procedure using IRAF software that comprises bias
subtraction, extraction of the spectra, and wavelength calibration. We 
used the standard spectroscopic stars  Feige~67 (Oke 1990), and BD+26 2606 (Oke
\& Gunn 1983) to  correct approximately for the response of the  configuration to different
wavelengths. Given the prohibitive time needed to obtain flat-field 
images (especially in the blue part of the spectrum), we did not correct for that effect. 
However, we have checked that this correction would have been very small ($\leq 1$ \%) and would not  have affected any of the  identifications of the main spectral features 
reported here. 
The spectrum of source~\#5 is shown in the top panel of  Figure~7. 
A strong MgII $\lambda$2798 \AA\ emission line and a weak emission line 
of O II $\lambda$3727 \AA\ are 
 present in this spectrum. 
From these emission-lines we determine that this object is a background quasar at a redshift of 1.365.

\begin{center}
\includegraphics[angle=90,width=\columnwidth]{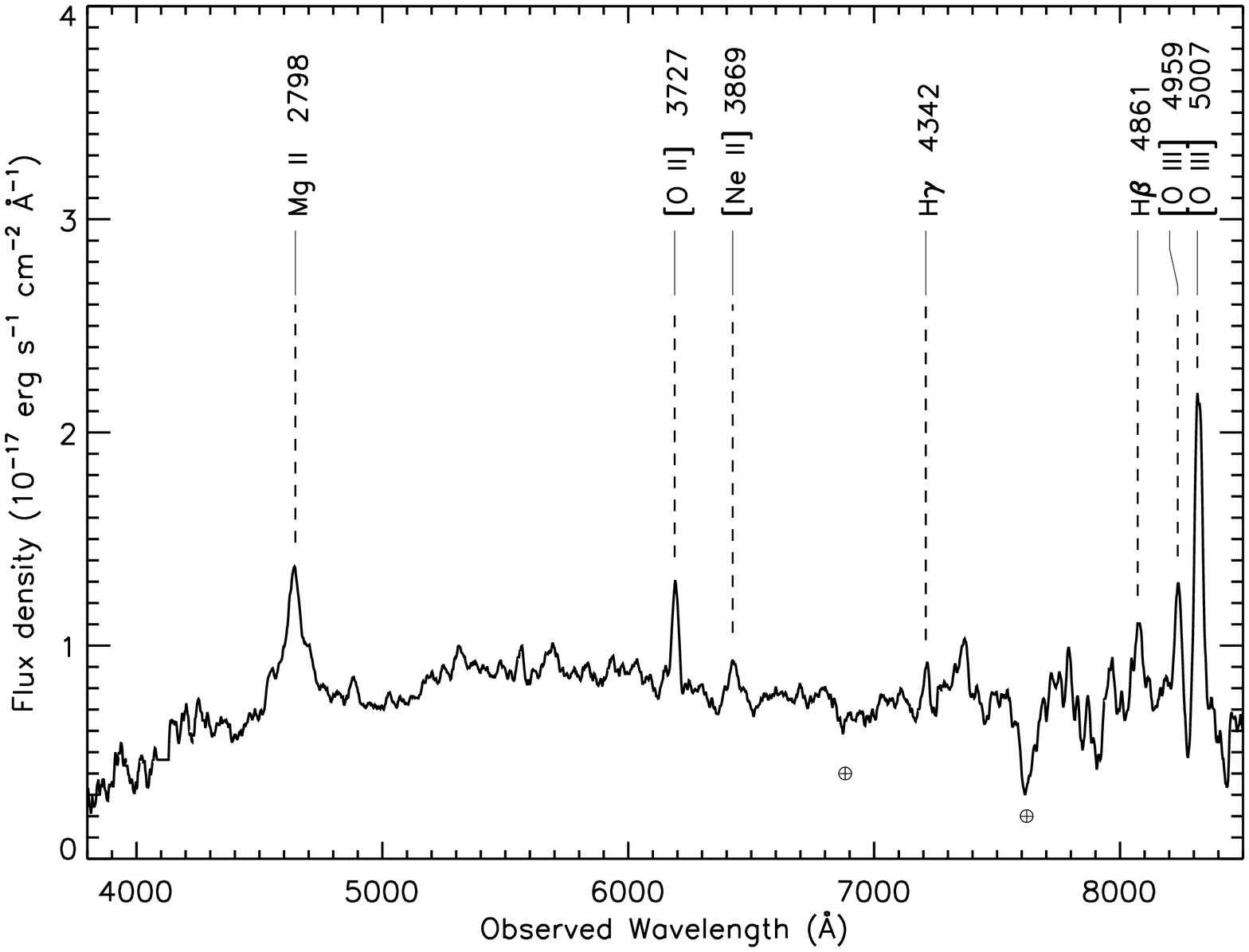}
\figcaption{Optical spectrum of source\#26, determined to be
a background quasar at a redshift of 0.6606. The circled crosses near $\lambda\lambda$6900 and 7200 
\AA\ denote the positions of strong telluric absorption in the Earth's 
atmosphere. The spectrum is smoothed with a boxcar of 11 pixels for 
display purposes.  }
\end{center}


The optical spectrum of source~\#26  was obtained with 
the 3.6~m Telescopio Nazionale Galileo (TNG) telescope on June 2005. 
A single exposure of 1800~s with a slit of 2\arcsec\ 
using the  DOLORES spectrograph and the MR-B grism  was taken.  For wavelength calibration we used He and Ar lamps. The spectrum was reduced according to standard procedure within IRAF as mentioned above. 
The spectrum is displayed in the bottom panel of Figure~7.
It is a background quasar at a redshift of 0.661 based on the 
 MgII $\lambda$2798 \AA\ emission line.

A second spectrum of source~\#26  was obtained at the ESO 3.6~m  
telescope using the ESO Faint Object Spectrograph  
(EFOSC2) on  June 6th, 2005. The grism Gr\#13 was used,  
providing a dispersion of $\sim$ 2.77 \AA\ pix$^{-1}$ with a wavelength  
coverage between 3720 and 9330 \AA. Two spectra of this source were obtained 
with an exposure time  
of 1800~s each and a signal-to-noise ratio of about 5-10. They were  
reduced using standard IRAF tasks, shifted to the heliocentric  
rest-frame and flux-calibrated with a spectrophotometric standard  
star. The spectral resolution is 440 at the central wavelength (6530  
\AA\, 5.4 pixel element). 
Some fringing is present at the level of $\sim$5-10~\% 
at wavelength long ward of 8000 \AA. The Mg II $\lambda$2798~\AA\
as well as Balmer lines H$\beta$ $\lambda$4861~\AA\ and H$\gamma$ 
$\lambda$4342~\AA\ are seen, as are forbidden lines of O and Ne. 
While Mg II is clearly broad, with a FWHM $\approx$ 5300 km s$^{-1}$, 
H$\beta$ is 
narrow, with a FWHM of only 400 km s$^{-1}$, 
similar to the forbidden line widths. However, most of the the narrow 
lines are either weak or affected by systematic effects such as 
fringing, and the typical error in the measurement of their width is 
about 40 per cent. This spectrum is shown in Figure~8.

The SDSS obtained optical spectra of the nuclear regions
 (circle of 3\arcsec\ radius corresponding to the size of the SDSS fiber) of 
 NGC~5774 and NGC~5775. 
The only strong emission lines present in the nuclear 
 spectrum of NGC~5774 are \ha, [N{\sc ii}] $\lambda$6583\AA,
 and [S{\sc ii}] $\lambda$6725\AA. 
The intensity ratios of these lines
 suggest that the emission is from normal H{\sc ii} regions (Ho et~al. 1993). 
The nuclear optical spectrum of NGC~5775 is much redder than that of 
 NGC~5774, as anticipated by the high inclination of NGC~5775, but 
 the spectrum is otherwise similar to that of NGC~5774 
 with relatively weaker emission lines.
No nuclear point source is detected in the Chandra image of either galaxy to a  
 limiting luminosity of 3$\times$10$^{37}$~ergs~s~$^{-1}$. 
Thus, we conclude that these galaxies do not host active nuclei.

\subsection{Infrared \& Radio Observations}

The only counterpart to an X-ray source detected in the infrared 
Two Micron All Sky Survey (2MASS) 
data is source \#5. Its $J$, $H$, and $K$ magnitudes
are 17.0$\pm$1.4 mag, 16.7$\pm$1.6 mag and 15.8$\pm$1.8 mag, respectively.
These colors are consistent with the quasar interpretation for this source (Cutri et al. 2002).

The NRAO Very Large Array~--~Faint Images of the Radio Sky at Twenty-centimeters (VLA--FIRST, Becker, White \& Helfand 1995) image of \ngc\ field is shown in Figure~9. 
Only source~\#47 has a potential counterpart in this image.
Its  radio peak flux-density is (2.94$\pm$0.15) mJy~beam$^{-1}$.  We also queried the positions of ULX candidates in the NRAO Very Large Array Sky Survey (NVSS and FIRST were operated at 1.4 GHz,  Condon, et al. 1998) and no  source was detected. This is expected, because the limiting peak source brightness of NVSS was about 2.5 mJy~beam$^{-1}$ (Condon, et al. 1998). 

Radio observations of the NGC 5774 / 5775 galaxy pair were carried out
using the Australia Telescope Compact Array (ATCA) in January 2005. The
continuum observations (over a time-slot of 7 hours) were made at all four
frequencies 1.4, 2.3, 4.8 and 8.4 GHz by alternately observing at 1.4/2.3
GHz and 4.8/8.4 GHz frequency pair using turret rotation. The array used
for the observations was 1.5D. The synthesized beams are elongated in
declination as a consequence of the northern declination of the source and
hence only the results for the higher frequency pair have been considered.
The data at the two frequencies were reduced using standard procedure
within the MIRIAD package (Sault, Teuben \& Write 1995; Sault  \& Killeen 1998).
The quoted fluxes are from the cleaned and
primary beam corrected images at the two frequencies. The rms noise
achieved at 4.8 GHz and 8.4 GHz was respectively 0.21 and 0.06 mJy and the
synthesized beams were 2.4" $\times$ 35" and 2" $\times$ 17". 
We have detected radio emissions from sources \#19, 47 and 49, with peak flux--densities of 1.58, 0.68 and 1.12 mJy beam$^{-1}$, respectively, at 4.8 GHz with a typical rms of 0.40 mJy. These results can be compared with results of the VLA--FIRST. However, it must be mentioned that the $ATCA$ and the VLA--FIRST measurements were made using two different instruments with different  calibration, imaging and analysis methods. In addition, 1.4 and 4.8 GHz frequencies were used for the VLA--FIRST and the $ATCA$ observations of \ngc\, respectively.
Thus, it is important that  we compare the peak flux densities at a common frequency, say at 1.4 GHz. To convert the $ATCA$ peak flux densities (S$_{\nu}$) from  4.8 GHz to 1.4 GHz, we need to know the value of radio spectral index. Results of radio observations of ULXs suggest that  the values of radio spectral indices ($\alpha$) are in the range of 0.4 to 1.2, where S$_{\nu}$~~$\propto$~$\nu^{-\alpha}$ (Neff, Ulvestad \& Campion 2003; Kaaret et al. 2003; Kording, Colbert \& Falcke 2005; Miller, Mushotzky \& Neff 2005; Soria et al. 2006; Lang et al. 2007). We have used the conservative-value of $\alpha$ as 0.5  and the estimated 1.4 GHz peak flux densities for sources \#19, 47 and 49 are 2.95$\pm$0.56, 1.24$\pm$0.56, and 2.08$\pm$0.56 mJy, respectively. Comparison of these results with that of the VLA--FIRST results suggest that the 1.4 GHz peak flux densities of these three sources varied by (2.95-0.15)$\pm$0.58, (2.94-1.24)$\pm$0.58 and (2.08-0.15)$\pm$0.58 mJy, respectively, where 0.15 mJy is the typical rms of the VLA--FIRST observations. Considering the large uncertainties, which were computed through the propagation of errors, we find that  sources\# 19 and 49 were variable at or above the three sigma level. Even though these two sources appear variable, but these results should be taken cautiously because of the differences between the VLA--FIRST and the $ATCA$ observations and analysis procedures. However, it should be emphasized that whether sources \#19, 47 and 49 are radio variable or not, but they are radio emitters.


\begin{center}
\includegraphics[angle=-90,width=\columnwidth]{Fig9.ps}
\figcaption{The VLA-FIRST image of \ngc\ with  D$_{25}$ isophotes. Three circles of radius 10\arcsec\ are also shown at the positions of three sources detected with ATCA.}
\end{center}


\section{Discussion}

Radio through X-ray observations of the interacting pair of galaxies, \ngc, have enabled a detail investigation of the dynamical, morphological and spatial properties of this system.

\subsection{The cumulative X-ray luminosity function}

There are eight (excluding the AGN, s05) X-ray sources in NGC~5774. Grimm, Gilfanov, \& Sunyaev (2003a) proposed a universal cumulative X-ray luminosity function (XLF) of high-mass X-ray binaries (HMXRBs), suggesting a  normalization factors (NF) proportional to the 
star formation rate (SFR) of a galaxy. Thus, the formalism of Grimm, Gilfanov, \& Sunyaev (2003a)
would predict only one HMXRB in NGC~5774 with 2--10 keV luminosity $\sim$1.78$\pm 0.3)\times10^{38}$~\ergl. This is extremely small compared to the observed total X-ray luminosity of the eight point sources of NGC~5774, which is (6.85$\pm 0.39)\times10^{39}$~\ergl\ (Table~1). Such huge discrepancy  may be due to the non-linear relation between the total luminosity of HMXRBs and  the low-SFR of NGC~5774 (Grimm, Gilfanov, \& Sunyaev 2003a).
 NGC~5774   has a low SFR of 0.2~\msun\ yr$^{-1}$, but the presence of three ULXs and five other X-ray binaries is remarkable.  Using the mass and the SFR of NGC~5774 and Equation (4) of Colbert \etal\ (2004) we can determine the total point-source X-ray luminosity of this galaxy, which is (4.15$\pm 0.5)\times10^{39}$~\ergl. This is comparable to the observed value and suggests that the X-ray source population of NGC~5774 consists of the older stellar population (including low mass X-ray binaries) and the younger stellar population (including HMXBs), which are directly related to the mass and SFR of NGC~5774 (Colbert \etal\ 2004; Grimm, Gilfanov, \& Sunyaev 2003b).
Three ULXs of NGC~5774 are located in active starforming regions on the spiral arms of this galaxy.  This indicates that the formation of ULXs  strongly depends on its immediate local environment but not on the global properties of the host galaxy.

Twenty four  X-ray sources (excluding the AGN, s26) are detected in NGC~5775. 
Observed counts of these twenty four sources are plotted against the cumulative number of sources in Fig. 10. 
It can be seen from this figure that the XLF of the starburst galaxy NGC~5775 is best represented as a single powerlaw (slope=$-0.71$) whereas the XLF of all the sources studied requires a broken powerlaw with a steeper slope at high flux. For NGC~5774, both a single and a broken powerlaw provide acceptable fits (there are too few sources to differentiate) but the slope of the single powerlaw is $-0.82$, steeper than that of NGC~5775.
The value of the slope of the XLF of NGC~5775 is consistent with the mean values of -0.65$\pm$0.16 for the interacting/merging and -0.79$\pm$0.24 for spirals (Colbert \etal\ 2004).
The XLFs of X-ray sources in 
 spiral galaxies with moderate star formation rates are relatively flat 
 (e.g., Kilgard \etal\ 2002;
 Colbert \etal\ 2004), which means that the most luminous individual sources 
 dominate the total X-ray luminosity (L$_{X}$) from these galaxies (Grimm, Gilfanov, \& Sunyaev 2003b).
Comparison of XLFs between NGC~5775 and interacting/merging galaxies shows that their slopes are also around -0.70 with a dispersion of 0.2, even though they have different star formation rates (SFRs) [M82 (Kaaret et al. 2001), NGC~253 (Strickland et al. 2000), NGC~3256 (Lira et al. 2002), NGC~3395/3396 (Brassington, Read \& Ponman 2005), NGC~4485/4490 (Roberts et al. 2002), The Mice (Read 2003), The Antennae (Fabbiano et al. 2001), The Cartwheel Ring Galaxy (Gao, Wang, Appleton \& Lucas 2003)].
However, it may be noted that the normalization factors (NFs) of respective XLFs are different for these galaxies.
Here we would like to mention that the number of sources detected  above 2$\times$10$^{38}$~\ergl\ in  NGC~5775, are consistent with the number of HMXRBs predicted by the universal XLF of Grimm, Gilfanov, \& Sunyaev (2003a). Thus, the XLFs with same  slopes but different  NFs indicate for evolutionary stages of galaxies. For example, comparison of XLFs between NGC~5775 and The Antennae or NGC4485/4490 shows that  these galaxies have similar spectral slopes but the NF values are large for The Antennae or NGC~4485/4490. These results suggest that The Antennae or NGC~4485/4490 are more evolved than that of NGC~5775. This is independently supported by the radio results, which suggest that the star formation in NGC~5775 was
 triggered by an encounter between the two galaxies some 100~Myr ago
 (Irwin 1994).

\begin{center}
\includegraphics[angle=-90,width=\columnwidth]{Fig18.ps}
\figcaption{The X-ray luminosity functions of  Chandra sources is shown for all sources in 6'.6 $\times$5'.8 field (solid histogram) and for sources in NGC~5775 (dotted histogram), NGC~5774 (dot-dashed histogram), and outside these galaxies (dashed histogram). Also shown is the best-fit single powerlaw model (slope -0.71) to the luminosity function of NGC~5775 and broken-powerlaw fit to all the sources.}
\end{center}

\subsection{Starburst activities}

SDSS obtained optical spectra of the nuclear regions
 (circle of 3\arcsec\ radius corresponding to the size of the SDSS fiber) of 
 NGC~5774 on May 16, 2001.
The only strong emission lines present in the nuclear 
 spectrum of NGC~5774 are \ha, [N{\sc ii}] $\lambda$6583\AA,
 and [S{\sc ii}] $\lambda$6725\AA. 
The intensity ratios of [N{\sc ii}] $\lambda$6583/\ha\ is less than 0.6. This suggests that
 the emission-lines are formed in the nuclear  H{\sc ii} regions (Veileus \& Osterbrock 1987; Ho et~al. 1993; 1997). 
No nuclear point source is detected in the Chandra image of this galaxy to a  
 limiting luminosity of 3$\times$10$^{37}$~ergs~s~$^{-1}$. Source \#~3 (s03) is close to the optical center, but the astrometric
corrected Chandra position of s03 is $\sim$10.\arcsec0 away from the optical nucleus.
Thus, we conclude that this galaxy does not host active nucleus to a  
 limiting luminosity of 3$\times$10$^{37}$~ergs~s~$^{-1}$.
SDSS  spectrum of the nuclear region of 
NGC~5775 is much redder than that of 
 NGC~5774, as anticipated by the high inclination of NGC~5775, but 
 the spectrum is otherwise similar to that of NGC~5774 
 with relatively weaker emission lines.
No nuclear point source is detected in the Chandra image of this galaxy to a  
 limiting luminosity of 3$\times$10$^{37}$~ergs~s~$^{-1}$. 
Thus, again we conclude that this galaxy also does not host active nucleus.

We extracted the X-ray spectrum of the diffuse emission from the D$_{25}$ region, excluding the point sources. The spectrum was binned to obtain at least 20 counts per fitting bin. A corresponding background spectrum was extracted from an ellipse of twice the D$_{25}$ area from a source-free region of the same CCD. Absorbed single powerlaw model does not fit well  with this spectrum ($N_{\rm H}$ consistent with the Galactic value and $\Gamma$= 3.2$^{+0.83}_{-0.37}$ for $\chi^{2}$=91.2 for 67 degrees of freedom). Similarly, absorbed single component thermal models like blackbody, mekal, bremstrahlung, disk blackbody, etc., fit poorly with this spectrum.
Finally, we find that absorbed $mekal$ plus powerlaw model fits well with the diffuse gas spectrum of NGC~5775 ($\chi^{2}$=65.2 for 65 degrees of freedom). 
The spectrum of diffuse gas, model and residuals are shown in Fig. 11. In this model 
$N_{\rm H}$ was required to freeze with the Galactic value. The $mekal$ temperature and abundance values are  $\sim$0.31$\pm$0.04~keV and 0.05$^{+0.20}_{-0.03}$, respectively. Photon index of the powerlaw component is 1.75$^{+0.68}_{-0.51}$. From the X-ray spectral fits of the diffuse gas, we have found that the 
fluxes  from the $mekal$ and the powerlaw components are 2.6$\pm$0.7 and 5.4$\pm$1.5$\times$10$^{-14}$~ergs~s$^{-1}$ in the 0.5--8.0~keV band, respectively. Parameters of the $mekal$  component can be used to compute different parameters (emission measure, particle number density, total gas mass, thermal energy, cooling time, etc.) to determine the physical properties of the diffuse gas. However, large uncertainties of the volume-filling factor will introduce huge errors in the values of these parameters (Strickland \& Stevens 2000). Instead, we will use the star formation rate to determine the origin of the diffuse gas.
 The temperature of the diffuse gas is slightly lower than  the expected diffuse gas temperature of $\sim$0.5$\pm$0.07 keV detected in interacting galaxies (Read \& Ponman 1998). This indicates that NGC~5775 is in early phase of interaction. The value of the abundance from the $mekal$ model suggest that the metallicity of the diffuse gas is sub-solar. 
Kewley, Geller, \& Barton (2005) have shown that the metallicity of gas
 in the central regions of starburst galaxies in pairs like \ngc\ tends
 to be sub-solar. They attribute this to channeling of metal-poor fuel 
 from the outskirts of the galaxies toward the central star-forming
 regions. Most likely, the NW and SE bridges are transporting outer-materials into NGC~5775 and  thus we expect the diffuse gas of this galaxy  to be composed of sub-solar materials.

We have detected outflow from the central region of NGC~5775, which is shown in Fig. 12. The observed outflow may be the starburst-driven superwinds from the starburst galaxy, NGC~5775. This indicates that NGC~5775 is in the superwind phase, which can be used as tool to study its evolutionary process.
Comparison of the observed results with the post-starburst evolutionary model results of Taniguchi et al. (2000) suggest that NGC~5775 have gone through both the starburst phase and the early-B star phase and entered into the superwind phase. This means that the wide-spread starformation took place around 2$\times$10$^{7}$~yr ago (Taniguchi et al. 2000). This is consistent with the estimated age of the interaction from dynamical considerations (Irwin 1994). In addition, these results are also consistent with the independent results derived from the studies of XLF and the X-ray spectrum of the diffuse gas of NGC~5775.

The fact that star formation is occurring throughout the disk of NGC~5775 yet there is very little visible distribution of the disk/spiral structure of the galaxy pair indicating that the starburst in NGC~5775
 is triggered by tidal forces redistributing the gas within the galaxy
 disk rather than by gas transfer between the galaxies.
Thus, the  hot diffuse gas was produced during the wide-spread starformation episodes in NGC~5775 and is proceeding at a rate of 5.4~\msun\ yr$^{-1}$.  Diffuse gas has two origins: thermal and non-thermal. Thermal component is due to supernova remnants (SNRs) and stellar winds. Contributions from SNRs can be computed using a relationship between the far infrared (FIR) luminosity and the supernova rate of starburst galaxies, which is as follows  (Mattila \& Meikle 2001):

\begin{equation}
r_{SN}=2.7\times10^{-12}~L_{FIR}/L_{\odot}~{yr}^{-1},
\end{equation}

\noindent The FIR luminosity (L$_{FIR}$) of NGC~5775 is 3.14$\times$10$^{10}$~L$_{\odot}$ (Devereux \& Eales 1989), which corresponds to 0.084 SN per year. Thus, $1.7\times10^{6}$ SNe exploded during the interval of 2$\times$10$^{7}$~yr.
If we assume that a massive star evolves to explode into a SN in 10$^{7}$~yr and 10$^{51}$~ergs is released from each SN explosion, then total thermal energy associated with the diffuse gas of NGC~5775 will be 1.7$\times$10$^{57}$~ergs. In addition, there will be some more contribution to the thermal energy from stellar winds. Of course, part of the total thermal energy will be lost due to radiative processes, gravitational potential energy and kinetic energy of the galactic wind/outflow. Large fraction of the thermal energy will be dissipated during the superwind phase of NGC~5775, which may last $\sim8\times10^{7}$~yr (Taniguchi et al. 2000).

Results of the spectral fitting show that the powerlaw component contributes more than double to the diffuse gas compared to the $mekal$ component. 
 Powerlaw sources are most likely unresolved X-ray binaries plus x-ray emitting stars. In addition, we find that the total diffuse gas contributes  less than $\sim$5.0\% to the total X-ray luminosity of NGC~5775. These results along with the non-detection of the nuclear source plus  detection of the galactic wind  suggest that NGC~5775 is in the beginning of the evolutionary process similar to NGC~3395/3396 (Brassington, Read \& Ponman 2005). 
This is further supported by the evolutionary model of  Taniguchi et al. (2000), which suggests that, presently, NGC~5775 is in the superwind phase
and after spending $\sim8\times10^{7}$~yr this galaxy will enter into the LINER phase (Taniguchi et al. 2000). However, it may be mentioned that the evolutionary stage of NGC~5775 is determined based on the evolutionary models of Taniguchi et al. (2000), which have been constructed to explain the optical narrow emission-lines. These models are yet to be examined with the multiwavelength properties of starburst galaxies.
Thus, presently, we are not certain about the evolutionary tracks of these models and the evolutionary results of NGC~5775, present here, should be taken cautiously.

\begin{center}
\includegraphics[angle=-90,width=\columnwidth]{diffuse_gas1.ps}
\figcaption{Upper panel shows the \cxo\ spectrum of diffuse gas in D$_{25}$ region of NGC~5775, which is fitted with the absorbed $mekal$ plus powerlaw model with  $N_{\rm H}$ fixed at the Galactic value, 3.47 x 10$^{20}$~cm$^{-2}$.  Lower panel shows the residuals between the data and the model.}
\end{center}

\begin{center}
\includegraphics[angle=-90,width=0.6\columnwidth]{HST_ACS_F658N.ps}
\figcaption{$HST/ACS/WFC$ \ha\ (F658N) image of NGC~5775. Galactic wind or ouflow from the central region of NGC~5775 can be seen at the south-west corner of the image.}
\end{center}

\subsection{Sources on the bridges and within the field}

From the comparison of Figs. 1 and 2, we find that sources \#14, 16 and 25 are located on the NW-bridge region. We know that source \#25 is associated with a background galaxy. Sources \#14 and 16 have no optical counterparts and are most likely  not background AGNs. Most likely, they are HMXRBs.
It appears that source \#16 is probably associated with the halo of NGC~5775. Thus, we conclude that  
source \#14 is an HMXRB located on the NW-bridge. Similarly, we find that sources \#6, 10, 15, 18, and 23 are located on the SE-bridge. 
Based on the optical colors,  sources~\#6 and 15 are possible background AGNs.  Source \#23 
 is a potential globular cluster (Fig.6 and section 2.2). No objects brighter than 25 mag in the $V$-band are present within the error circles at the astrometric corrected \cxo\ positions of sources \#10 and 18. Both the sources \#18, and 23 are located very close to the D$_{25}$ isophote of the edge-on galaxy, NGC~5775. Thus, we suggest that they are most likely the halo-objects of NGC~5775. 
Finally, we are left with two sources - s10 and s14, which are  around $2\times10^{37}$~ergs~s~$^{-1}$ and are probably HMXRBs formed on the SE- and NW-bridges, respectively. Presence of these objects provide evidence for recent starformation on the bridges that might have been initiated by the interaction between the galaxy pair (Smith et al. 2007). Sources \#18, and 23  are ULX candidates.
\cxo\ spectra of these two ULX candidates can be described by 
 the powerlaw model (Table~1). The X-ray to optical flux ratio (\fxfo) is larger than 13 for source~\#18 and source~\#23 is hosted in a globular cluster.

Sources \#2, 13, 17, 19, 29, 39 and 48  are within the field but they are located neither on the bridges nor within the D$_{25}$ isophotes of two galaxies. Properties of sources \#2, 13  and 39 are discussed in section 2.2. Source \#17 is associated with a  blue compact dwarf galaxy (Fig. 5), which contains large clusters of young, hot and massive stars . Radio emission was detected from source \#19 and most likely this is a B L Lac object. We could not detect the optical counterparts of source \#29 and 48 and they are close to D$_{25}$ isophote of NGC~5775 and they are most likely HMXRBs in the halo of this galaxy. However, it should be mentioned that hydrodynamical simulations of interacting galaxies suggest that the gas in the connecting bridges is shocked or compressed due to interaction, which triggers wide-spread starformation
episode and some gas may fall onto the outer disk  (Struck \& Smith 2003). If this is true, then we would expect many X-ray sources to form at the interface of the disk and bridges.  Thus, the halo-objects of NGC~5775 (s16, s18 and s23), which are also located at the interface of the disk and bridges, might have formed due to the interaction between the outer disk and the falling gas from the bridges.

\subsection{Nature of the ULXs}

In this section, we describe interesting individual sources in some detail, particularly the most luminous X-ray sources which are candidate ULXs.  
In total, forty-nine sources were detected in two X-ray observations within the 6.\arcmin 6$\times$5.\arcmin 8 field (Figs. 1 and 2)
above a detection limit of $\sim$10$^{38}$~\ergl.
Nine X-ray sources were detected near or within  
 the D$_{25}$ isophote of NGC~5774.
There are 25 X-ray sources within the D$_{25}$ isophote of NGC~5775. 
Rest fifteen sources were also detected within the field but they were  out side the D$_{25}$ isophotes of these galaxies.
Properties of these fifteen objects have  already been discussed in sections 2.2 and 3.3. Here we will discuss about the candidate ULXs
detected within the D$_{25}$ isophotes of NGC~5774 and  NGC~5775.
Due to the large amounts of unresolved optical emission from the disk of
NGC~5775 and its nearly edge-on orientation, identification of potential
optical counterparts to the detected X-ray sources cannot be made for all the sources with high
confidence. Thus, we concentrate mainly on the X-ray properties of the sources
in most cases. 

In 1996, a bright supernova was detected in NGC~5775, SN~1996ae 
(Vagnozzi  et al. 1996).
We used the Chandra and XMM-Newtom data to search for X-ray emissions from SN~1996ae. No source was detected at the location of this supernova. A two sigma upper limit of 2 x 10$^{37}$ ~ergs~s$^{-1}$ was determined.

Before we switch over to candidate ULXs, we have to determine the possible foreground/background objects, which mimic as ULXs.
 Since the value of \fxfo\ for foreground stars will be less than 0.1 and that for AGNs is of order unity 
(e.g., Green \etal\ 2004), the brightest X-ray sources 
with optically bright counterparts are very likely either Galactic stars or high-redshift AGNs.
The flux from a ULX ($L_{\rm X}>10^{39}$~\ergl)
at the distance of NGC~5774/5775 is
8.4$\times$10$^{-15}$~ergs~cm$^{-2}$~s$^{-1}$ in the 
2--10~keV bandpass assuming a $\Gamma=1.8$ power law spectrum
and an absorption column density equal to the Galactic value.
We expect 3 background sources at or above this flux level 
within the 6.\arcmin 6$\times$5.\arcmin 8 field analyzed here,
according to the log($N$)--log($S$) relation 
deduced from the \cxo\ Deep Field South survey (Rosati \etal\ 2002).
Two ULX candidates with bright optical counterparts
lying at or just beyond the D$_{25}$ isophotes of the NGC~5774/5775
galaxy pair have spectroscopically-confirmed redshifts higher
than that to the galaxy pair.
Discovery of high redshifts to ULX candidates outside the 
optical extent of target galaxies is now a common occurence
(e.g., Arp, Guti\'{e}rrez,   \& L\'{o}pez-Corredoira 2004; Guti\'{e}rrez \& L\'{o}pez-Corredoira 2005; Guti\'{e}rrez  2006)
and agrees with the statistical evidence that 
the radial distribution of ULXs in galaxies asymptotically
approaches the expected background level at roughly the radius of the 
D$_{25}$ isophote (Swartz \etal\ 2004; Swartz  2006). Based on the SDSS-colors, we have identified s06, s09, s13 and s15 are likely background AGNs. The two ULX candidates, s18 and s23, have already been discussed in section 3.3 and  will not be repeated here.
Finally, we are left with 
three and seven candidate ULXs  in NGC~5774 and NGC~5775, respectively. These ten objects will be  described in the following section.
SDSS images of individual sources, their Chandra spectra and spectral results are presented in Fig. 5, Fig. 3 and Table~1, respectively.

\subsubsection{Source~\#1}
The Chandra light curve of source~\#~1 displayed significant variability. Its X-ray light curve is shown in Figure~13. Assuming a constant count rate 
 results in a value of $\chi^{2}$ of 85.7 for 58 degrees of freedom and a KS 
 probability $<$ 10$^{-5}$. 
The Chandra spectrum of this ULX candidate can be 
 described as an absorbed powerlaw. 
This source was not detected during the XMM--Newton  
 observations. We queried at the astrometric corrected Chandra position of this source and the derived value of L$_{X}$ is 
$\sim~7.7\times10^{38}$~ergs~s~$^{-1}$ in the 0.5-8 keV band. Comparison of this value with that of Chandra, listed in Table~1, shows that it varied by a factor of three between these two observations separated by around 15 months.

HST/WFPC2 image in F814W filter was used to search for the optical counterpart of source~\#~1. The astrometry between the Chandra and the HST images were performed using the USNO-B1.0 star 0935-0243054 and the estimated astrometric error is $\sim$0.\arcsec70.  
No optical counterpart was detected within the error circle at the astrometrically-corrected Chandra position of the X-ray source, to the limiting  
 magnitude of $\sim$26 mag. 
The source is located on the spiral arm and also within a star forming complex. The \hst\ image shows many bright knots near this ULX candidate that are
 also apparent in the SDSS composite image.

Despite the lack of an optical counterpart, this source is similar in the X-ray temporal and spectral 
 behavior to the transient ULX in Cen~A that has been attributed to a
 Be/X-ray binary (Ghosh \etal\ 2006). 
At the distance to NGC~5774, a late O or early B companion to source \#1 
 would not be detectable even with \hst.

\begin{center}
\includegraphics[angle=-90,width=\columnwidth]{Fig14.ps}
\figcaption{\cxo\ light curve of source~\#~1 with 1000~s binning.}
\end{center}


\subsubsection{Source~\#4}

This source was  detected with \cxo\ but was absent  at the 2$\sigma$ detection limit of 
$\sim$3~$\times$ 10$^{38}$ ~ergs~s$^{-1}$ during the XMM-Newton/PN observation. This source falls near the Chandra CCD gap, which suggest that it was probably more luminous during the Chandra observation than that listed in Table~1.
Its \cxo\ spectrum is  described with the powerlaw model. KS-test probability of $\sim$0.001 indicates it was  variable during
the \cxo\ observation.  Source~\#4 is located at the inner portion of the inner 
spiral arm of NGC~5774. Close inspection of the SDSS image shows the presence of uniformly distributed diffuse emission at and around the astrometric-corrected \cxo\ position of this ULX candidate. 
The ESO/NTT \ha\ image shows that this source is located at an elongated diffuse emission region.  
It being a X-ray variable object, we suggest that the diffuse emission is a nebula formed by the injection of kinetic energy by a jet from this ULX.

\subsubsection{Source~\#12}

This source lies just beyond the D$_{25}$ isophote of NGC~5774 but 
 is located on the spiral arm of this galaxy (see Figure~1). 
Eventhough this source was not detected with \xmm\, but the extracted source countrate suggest that it is not a transient X-ray source. 
It is not a highly absorbed  source and its Chandra spectrum suggest that it may be a flat spectrum object. No optical counterpart was detected in the ESO/NTT \ha, R-band and SDSS images.

\subsubsection{Source~\#24}

Source~\#~24  was bright (L$_{X}$ $\sim~7.78\pm0.32~\times10^{39}$~ergs~s~$^{-1}$ in the 0.5-8 keV band) during the \cxo\ observation but not detected in the \xmm\ data with a 2$\sigma$ detection upper limit of 5.6$\times$10$^{38}$~\ergl\ in the 0.2-10~keV band. This shows that it varied by more than a factor of 10 between the two observations i.e. over a period of 15 months. However, it was steady during the interval of \cxo\ observation. Chandra spectrum of this source fitted with an absorbed powerlaw model is shown in Figure 3 and the spectral parameters are listed in Table~1. Residuals between the data and the model are shown in the lower panel of Figure 3. It can be seen from the residual plot that an emission feature around 3 keV is present in the spectrum of this absorbed and flat powerlaw object. We  added a Gaussian emission-line component to the powerlaw model to fit the Chandra spectrum. From the fit parameters, we find that $\Delta\chi^{2}$=7.2 for 18 d.o.f and the F-test statistic and probability values are 5.92 and 0.0054, respectively, which suggest that the detection of the line at 2.99~keV (width=0.16~keV and equivalent width is 541 eV) is significant at $\sim$99.5\% level. It is important to mention that it is a flat powerlaw object ($\Gamma$=1.26$^{+0.60}_{-0.30}$). 
Figure~14 shows the $HST/ACS/WFC$ \ha\ (F658N) image of the region surrounding this ULX candidate. 
A  circle of 0.\arcsec2 radius, which is the size of the  error circle, is also shown. It may be seen that this source is located within the diffuse \ha\ emission region and close ($\sim$39~pc) to an object of absolute magnitude around -12.2~mag in the F625W filter. This object may be a young star cluster, because most of the H{\sc ii} regions are much brighter than this object (Lee et al. 2001). In fact, the  error circle covers part of this star cluster, suggesting that this ULX candidate is associated with this star cluster.

Flat powerlaws have been detected in the low/hard state spectra of Galactic black hole X-ray binaries (McClintock \& Remillard 2006). In this state, the accretion rate is sub-Eddington and compact radio jets may be present. In addition,  radio and X-ray intensities are correlated (McClintock \& Remillard 2006;  Remillard \& McClintock 2006). Thus, it is believed that X-ray beaming is absent and X-ray emission-lines can be detected in this low/hard state (Remillard \& McClintock 2006). Flat powerlaw and the presence of a broad ($\sim$0.16 keV) emission line at 2.99~keV (Ca~XX) in the  \cxo\ spectrum of this ULX suggest that beaming was not present in source \#24  and it was in low/hard state during the \cxo\ observations. If we assume that its observed X-ray luminosity was due to isotropic emission at the sub-Eddington ($\sim$0.02$\times$L$_{Edd}$) rate, then the mass of the accretor is $\sim$3000~\msun. It entered into further lower flux level during the  \xmm\ observations.  Such dramatic variation indicates that the accretion disk temperature at the outer edge was below the hydrogen ionization temperature. This can happen when the mass transfer rate falls below a critical value and the mass of the accretor will be much higher than those of stellar-mass black holes, if the companion is  a high-mass star (King et al. 1996). Source \#24 being located in the star forming region and most likely hosted in a young star cluster, it is expected that its companion will be a high-mass star. Thus,  source \#24 is a prospective intermediate mass black hole (IMBH) candidate.

\begin{center}
\includegraphics[angle=-90,width=0.6\columnwidth]{hst_acs_src24.ps}
\figcaption{$HST/ACS/WFC$ \ha\ (F658N) image around the astrometric-corrected \cxo\ position of  source~\#24. A circle of radius 0."2 marks the position of the X-ray source.}
\end{center}


\subsubsection{Source~\#28 }
Source~\#~28  was detected during the \cxo\ observation but was below the detection level in the \xmm\ data. Chandra spectrum of this source displays a narrow emission feature around 1.4 keV 
(Figure~3). 
This feature causes a poor fit statistic  
 ($\chi^{2}$=15.5 for 8 degrees of freedom) for source~\#~28 using a 
 powerlaw-only model. 
Adding a Gaussian emission-line model improves $\chi^{2}$ by 9.8 (F-test value is 5.22). 
The resulting 
 equivalent width of this line is 534~$\pm$~200~eV and the detection 
 of this line is significant at the 95.1\% level. The line center is 1.44$\pm$0.08~eV, which is most likely due to Mg~XII with an outflow velocity of around 1500~km~s$^{-1}$. No optical counterpart was detected within the error circle of radius 0.\arcsec2 at the 
astrometric corrected \cxo\ position of this source.

\subsubsection{Source~\#40}

It is  the brightest ULX candidate (7.6$\pm$0.7 $\times$ 10$^{40}$~ergs~s~$^{-1}$ in the 0.5-8~keV band) among 82 nearby galaxies in the survey of ULXs  described in Swartz et al. (2004). It was not detected by \xmm\ with a 2$\sigma$ detection upper limit of 3.5$\times$10$^{38}$~\ergl.
The 2--10 keV flux did not vary, 
but the 0.2 - 2 keV light curve, shown in Figure~15, indicates for possible variability during the \cxo\ observation ($\chi^{2}$/dof~$\sim$75/58).  This type of variability was not detected in any of the sources of Table~1 that had  at least 100 counts in total, compared to 1354 counts in source~\#40.
We have searched for pulsations in the soft, hard and total light curves  of this source, but no dominant pulsation was detected. Close inspection of the soft lightcurve suggest that this source can be a periodic variable, which is not clearly visible due to the presence of flares. We have fitted this lightcurve with a periodic variable component ($\sim$6.2~hrs period) plus flares, which could be from the accretion disk (Fig. 15). 

\begin{center}
\includegraphics[angle=-90,width=\columnwidth]{Fig11.ps}
\figcaption{The \cxo\ soft--X-ray (0.2--2.0) keV light curve of source~\#40 with 1000~s binning.
This light curve suggests weak variability ($\chi^{2}$/dof~$\sim$75/58). This variability is absent in the 2--10 keV light curve. If this energy-dependent variability in the soft band is periodic, it suggest that the system is either a wind-fed binary or the optical companion has an atmosphere, which absorbs the soft X-rays
during the observation.}
\end{center}

Both absorbed powerlaw and disk blackbody models fit equally well with the Chandra spectrum of source~\#40. Powerlaw spectral parameters are listed in Table~1. Disk blackbody model parameters are: T$_{in}$=1.82$^{+0.36}_{-0.20}$~keV, Normalization =0.00304$\pm$0.0015  and $\chi^{2}$/dof = 109/110.  This model fits slightly better than the powerlaw model ($\Delta\chi^{2}$=3.9). Parameters  of the disk blackbody model suggest that the mass of the accretor of source~\#40 is  30$\pm$9.0~\msun\, assuming that the inclination angle ranges between 0 and 45$^{o}$.
$HST/ACS/WFC$ \ha\ (F658N) image of this ULX candidate field is shown in Figure~16.
 The astrometric-corrected \cxo\ position of this ULX is near to a starforming region (Lee et al. 2001), indicating that this source is not hosted in any star cluster.

\begin{center}
\includegraphics[angle=-90,width=0.6\columnwidth]{hst_acs_src40.ps}
\figcaption{$HST/ACS/WFC$ \ha\ (F658N) image around the astrometric-corrected \cxo\ position of  source~\#40. A circle of radius 0."2 marks the position of the X-ray source.}
\end{center}


Source \#40 in NGC~5775 is the most luminous ULX in our catalog (Swartz et al. 2004).
It displayed remarkable variability by more than a factor of 500 between \cxo\ and \xmm\ observations. 
In addition, its energy dependent light curve displays  variations with a possible period of 6.2~hr and the X-ray spectrum
suggest that the mass of the accretor is in the range of 20--40~\msun. Based on these results and using the formalism
described in Ghosh et al. (2006), we suggest  that this system may be similar to that of Cygnus~X3 (Paerels et al. 2000). 
The X-ray modulation in Cygnus X-3  is explained 
as variable scattering in the dense photoionized wind from a Wolf-Rayet companion (Paerels et al. 2000). Energy dependent light curve of 
of source~\#40 (Fig. 15) also indicates that the possible modulation may be either due to high absorbing column or due to occultation of the X-ray source by the mass transfer stream (White \& Swank 1982). If the former is true, then this ULX is most likely a wind-fed system. Otherwise, the mass transfer stream will collide with the accretion disk and that will generate strong turbulence. This will cause the disk to swell in the vicinity of the confluence and this bulge or thickened region of the accretion disk may cause modulation. In this scenario,  the dips will have variable depths. Long-term changes and instabilities in the outer structure of the disk will result in phase jitter and anomalous dips (White \& Holt 1982). Presence of flare-like features in the light curve of source \#40 may be due to instabilities on the disk. These can be tested from future observations over a longer period.

Intrinsic X-ray luminosity of source \#40 is $\sim$10$^{41}$~\ergl\ in the 0.2--10 keV energy band, which suggest that the mass of the accretor has to be at least 770~\msun\ even if it is accreting at the Eddington rate. However, from the $HST/ACS$ data we found that this ULX is not hosted in a star cluster, which can  host the IMBH. In addition, we have determined that the mass of the accretor of this ULX system will be in the range of 20-40~\msun\ and its the huge X-ray flux variability between the  \cxo\ and \xmm\ observations
could be due to either geometrical or relativistic beaming along our line of sight. High mass accretion rates lead to thick accretion disks causing geometrically beamed emission (Fabrika \& Mescheryakov 2001; King 2001; Fabrika 2004; Poutanen et al. 2007). Geometrical models suggest that the X-rays will be beamed through the geometrical funneling and will be absorbed by the H- and He-like ions of abundant heavy elements present in the photosphere of the funnel's inner wall. These blue-shifted absorption lines will be superimposed on the emergent multicolor-continuum spectrum from the funnel (Fabrika \& Mescheryakov 2001; Fabrika 2004; Poutanen et al. 2007). However, we could not detect any such line in the Chandra spectrum of this ULX, though the 0.5-8~keV continuum fits well with the multicolor disk black body model. Thus, it is very likely that this ULX is beaming relativistically. Relativistic beaming could be  either due to inverse-Compton or synchrotron mechanism.
In the inverse-Compton scenario, if the optical photons from the companion star are up-scattered by the jet, then the X-ray emission can power the ULX. X-rays produced by this process will have a broken-powerlaw shape with a break-energy at around 1~keV (Georganopoulos, Aharonian \& Kirk 2002). However, we do not find the signature of broken-powerlaw in the \cxo\ spectrum of this ULX. 
Relativistically beamed  synchrotron jet will produce spectral curvature ($\Gamma$/dE is $>$0), which was suggested based on  the EXOSAT spectra of X-ray selected BL Lac objects (Ghosh \& Soundararajaperumal 1995). Recently, this has been confirmed with high signal-to-noise ratio \xmm\ spectra of these objects. (Perlman et al. 2005). Thus, to search for the spectral curvature we divided the \cxo\ spectrum of source \#40 in the 0.5-2.0, 2.0-4.0 and 4.0-8.0 keV bands and fitted these spectra with the absorbed single-powerlaw model. Fig. 17 shows the plot of $\Gamma$ versus energy. A constant model was fitted to this plot that resulted with $\Gamma$=1.6 and $\chi_{r}^{2}$=1.64 for 2 dof. However, a powerlaw model with a spectral slope of 0.75 fits this plot with  $\chi_{r}^{2}$=0.038 for 1 dof. Using these results we find that the F-test statistic and probability values are 85.1 and 0.0687, respectively. This means that the powerlaw model is significant at 93.1\% level with respect to the constant model. These results are not robust, which will firmly establish that  d$\Gamma$/dE is $>$0. Thus, we consider these results as a possible signature for the spectral curvature.
Spectral curvature can also be measured using a $logarithimic parabola$ model (Perlman et al. 2005 and references therein):

\begin{equation}
dN/dE = ke^{-\sigma(E)N_{H, Gal}}
e^{-\sigma(E)N_{H, int}(1+z)} ~E^{(-\Gamma+\beta Log(E))}
\end{equation}

\noindent where N$_{H, Gal}$,  N$_{H, int}$ are the Galactic and intrinsic absorbing columns and  $\beta$ is the curvature parameter.
We fitted the \cxo\ spectrum of source \#40 using this model and determined the value of $\beta$ as 0.4$\pm$0.2.
Again, this is not a highly significant result, but indicates that this spectrum may be curved. 
If the observed flux variations of source \#40 are due to the presence and absence of jet emission, then we can assume that beaming has boosted X-ray emissions at least by a factor of 500. This can be equated with $\delta^{3+2\alpha_{x}}$, where $\delta$ is the Doppler boosting factor for the approaching jet with X-ray spectral index $\alpha_{x}$. This gives the value of $\delta$ as 3.5, which is smaller than the value of $\delta$ determined for the brightest ULX of NGC~5408 (Kaaret et al. 2003). 
Thus, we suggest that, most likely, the luminous X-rays from the source \#40 is due to relativistically beamed  synchrotron jet (Kording, Falcke \& Markoff 2002).

\begin{center}
\includegraphics[angle=-90,width=\columnwidth]{d_gamma_dE.ps}
\figcaption{Plot of photon indices versus photon energy of source \#40. The solid line represents the powerlaw fit with spectral slope of 0.75 and $\chi_{r}^{2}$=0.038 for 1 dof. The powerlaw model is significant at 93.1\% level with respect to the constant model. This indicates that the \cxo\ spectrum of this ULX may be curved, i.e. steeper spectra at higher energies with curvature that remains constant.}
\end{center}

\subsubsection{Source~\#41}

This ULX candidate was detected in both \cxo\ and \xmm\ 
observations. The X-ray spectra fits well with the powerlaw model. Photon index and the hydrogen absorbing column density were almost steady, within the error limits, between the two observations although the  0.5~--~8.0 keV flux 
varied by almost a factor of two. It is interesting to note that both the \cxo\ ($\Gamma=1.48^{+0.32}_{-0.24}$) and \xmm\ ($\Gamma=1.40^{+0.17}_{-0.14}$) spectral indices of s41 suggest that it is a flat X-ray spectrum source. If flat spectral indices are the signatures of the sub-Eddington accretion rate ($\sim$0.02$\times$L$_{Edd}$) at the low-hard state of this ULX, then its observed \xmm\ luminosity will correspond to $\sim$6.3$\times$10$^{41}$~\ergl\ as the intrinsic luminosity in the 0.2--10 keV energy band (McClintock \& Remillard 2006; Remillard \& McClintock 2006). This suggests that the mass of the accretor has to be at least 4846~\msun\ even if it is accreting at the Eddington rate and emitting X-rays isotropically.

$HST/ACS/WFC$ \ha\ (F658N) image of this ULX candidate field is shown in Figure~18.
 It can be seen from this figure that the source~\#41 is located within a bright \ha\ (F658N) complex. With the available data, we cannot determine whether this source is associated with a H{\sc ii} region or a star cluster.

\begin{center}
\includegraphics[angle=-90,width=0.6\columnwidth]{hst_acs_src41.ps}
\figcaption{$HST/ACS/WFC$ \ha\ (F658N) image around the astrometric-corrected \cxo\ position of  source~\#41. A circle of radius 0."2 marks the position of the X-ray source.}
\end{center}


\subsubsection{Sources~\#46 \& 47}
These two ULXs lie within 4\arcsec\ of each other and cannot be resolved by
\xmm. The sum of the luminosities of the two sources in the \cxo\ observation 
is within errors equal to the luminosity of the combined object in the \xmm\
observation. 
Thus, it is likely that both the sources did not vary between the \cxo\ and \xmm\ observations. 
Source \#46 is a relatively absorbed steep X-ray spectrum ULX.
$HST/ACS/WFC$ \ha\ (F658N) image shows that source \#46 is embedded in diffuse emission (Figure~19).
Radio emissions were detected in the VLA-FIRST and in our ATCA observations
from the region but the angular resolution is not sufficient to determine
which source is the likely radio emitter. 
However, astrometric results between the VLA-FIRST and the Chandra images of NGC~5775 suggest that the VLA-FIRST position is more consistent with s47. We have already discussed in section 2.4 that s47 did not vary between the  VLA-FIRST and the ATCA observations but is a strong radio emitter.
No optical counterparts were detected within the \cxo\
error circles in the $HST/ACS/WFC$ \ha\ (F658N) image around s47 but this source 
 is located $\sim$2\arcsec.0 away from  a bright-blue object (Figure~20). 
Photometric colors and absolute magnitudes, assuming that this object is in 
NGC~5775, suggest that it is a large star-forming region. Both the \cxo\ and the \xmm\ results clearly show that s47 is a flat X-ray spectrum ULX. Thus, the radio, optical and X-ray properties of this ULX is interesting. For example, the lack of any optical point source at the astrometric-corrected Chandra position of s47 suggests that it is not a background object because its value of \fxfo\ will be much larger than the upper limits of B L Lac object (REF). On the other hand, its X-ray results 
clearly show that this ULX was in the low-hard state and the observed high X-ray luminosity indicates for an accretor of mass more than        , if accreting at the Eddington rate and emitting isotropically (McClintock \& Remillard 2006; Remillard \& McClintock 2006). This has been discussed in section      .
Again, the scenario of a high-mass accretor is further supported by the radio/X-ray results, even though these observations were not simultaneous. 
 A ``fundamental plane" relating black hole mass (M), X-ray ($L_{\rm X}$) and radio luminosities ($L_{\rm R}$) 
in unbeamed  sources has been discovered (Corbel et al. 2000, 2003; Gallo et al. 2003; Falcke et al. 2004; Merloni, Heinz \& Di Matteo 2003, 2005). Using this  relation:

\begin{equation}
\log L_{\rm R}=(0.60^{+0.11}_{-0.11}) \log L_{\rm X}
+(0.78^{+0.11}_{-0.09}) \log M + 7.33^{+4.05}_{-4.07}.
\end{equation}

 \noindent and the observed radio/X-ray fluxes, we find that the mass of the accretor will be very high. However, large vertical spread ($\sigma_{\rm R}=0.88$) of this relation suggest that the errors will also be large. This puts a limit on the mass the accretor  between several hundreds to several thousands of solar-mass. Thus, based on the non-simultaneous radio/X-ray results we can not firmly claim that s47 is an IMBH system but it is a very promising candidate.

\begin{center}
\includegraphics[angle=-90,width=0.6\columnwidth]{hst_acs_src46.ps}
\figcaption{$HST/ACS/WFC$ \ha\ (F658N) image around the astrometric-corrected \cxo\ position of  source~\#46. A circle of radius 0.``2 marks the position of the X-ray source.}
\end{center}


\begin{center}
\includegraphics[angle=-90,width=0.6\columnwidth]{hst_acs_src47.ps}
\figcaption{$HST/ACS/WFC$ \ha\ (F658N) image around the astrometric-corrected \cxo\ position of  source~\#47. A circle of radius 0."2 marks the position of the X-ray source.}
\end{center}


\subsubsection{Source~\#49}
This  source was detected only in the \xmm\ data. Its flux remained steady during this observation. The X-ray spectrum of this ULX candidate fits well with the powerlaw model. 
No source was detected within the error circle at the astrometric corrected \cxo\ position of this source in the $HST/ACS/WFC$ F625W image. However, it appears from the F658N image (Fig. 21) that this ULX is located either within a spherical nebula or within a dissolved star cluster (Pellerin et al. 2007). The UVW2 image obtained from the XMM-Newton optical monitor's telescope shows that the ULX is embedded in the diffuse UV emission, which is the typical characteristic of dissolved star clusters (Pellerin et al. 2007). We have discussed in section 2.4 that s49 is most likely a radio variable ULX, which was 
detected  during our ATCA observations (1.12$\pm$0.1~ mJy~beam$^{-1}$ at 4.8 GHz). Using the ``fundamental plane" relation, equation (3), and the observed radio
and X-ray fluxes we estimate that the mass of the accretor as   8.3$\times$10$^{4}$~\msun.  Due to the large vertical spread ($\sigma_{\rm R}=0.88$) of this relation
we find that the mass of the accretor could as low  as a few hundred solar-mass. In addition,  the radio
and the X-ray observations were not simultaneous. Thus, we should consider these results cautiously. However, it should be mentioned that s49 is a potential IMBH candidate and future simultaneous radio/X-ray observations are highly essential.

\begin{center}
\includegraphics[angle=-90,width=0.6\columnwidth]{hst_acs_src49.ps}
\figcaption{$HST/ACS/WFC$ \ha\ (F658N) image around the astrometric-corrected \cxo\ position of  source~\#49. A circle of radius 0."2 marks the position of the X-ray source.}
\end{center}


Ten ULXs are detected within the D$_{25}$ isophotes of NGC~5774 and NGC~5775. 
Sources~\#1, 12, 28 and 40 are all located within bright starforming complexes and no optical counterparts have been detected within the astrometric corrected Chandra positions of  these sources. Based on their observed properties we suggest that these are stellar-mass black hole systems.

X-ray spectra of sources~\#4 and 46 are steep and they are embedded in diffuse emission, which may be ionized nebulae. Ionized nebulae with bubble-like morphology are detected around some ULXs (Roberts et al. 2003;
Pakull \& Mirioni 2003; Pakull, Grise \& Motch 2006). These bubbles are several hundred parsec in diameter with expansion velocities around 50--80 km~s$^{-1}$ (Rosado et al. 1981, Rosado et al. 1982, Valdez-Guti\'errez et al. 2001).  It has been suggested that the  expansion could be due to the combined action of energetic supernova explosions and stellar winds, 
or to continuous inflation by  jets (Miller 1995,   Valdez-Guti\'errez et al. 2001, Wang 2002, 
Pakull \& Mirioni 2003, Pakull, Grise \& Motch 2006). Some nebulae show barrel-type shapes or enhanced 
emission along  opposite directions that could be interpreted as excitation from a beamed source (Roberts et al. 2003;
Pakull \& Mirioni 2003). However, some nebulae have displayed spur-shape, which requires an isotropic flux of energetic photons  to explain the observed HeII flux (Kaaret et al. 2004; Pakull, Grise \& Motch 2006). 
Nebulae formed by the combined action of supernova remnants and stellar winds are not supposed to have HeII emission unless it contains very massive stars.
Thus, detection of an
non-spherical HeII nebula at the position of the ULX will support the beaming model of these two ULXs. Nebulae formed around IMBHs will be much smaller in size compared to the nebulae formed by SNe or jets. Thus, the observed sizes of diffuse emission around s04 and s46 suggest that they are not IMBH systems.

Another mechanism that is prevalent in interacting galaxies and 
 may lead to the formation of luminous X-ray sources is the formation
 of super star clusters (SSCs).
There is a general correlation between the formation of SSCs and
 galaxy mergers (Whitmore 2000).
Keel \& Borne (2003) have shown that this trend applies also to
 systems like \ngc\ and does not require direct contact to occur.
Theory suggests that the most massive black holes may come from mergers of 
 massive main sequence (Portegies~Zwart \etal\ 2004) or 
 protostars (Soria 2006) within these young massive SSCs. 
If these intermediate-mass black holes with masses $\gg$20~\msun\ 
 can form in SSCs and can capture a stellar companion (Baumgardt \etal\ 2005)
 then X-ray luminosities near the Eddington limit, 
 1.3$\times$10$^{38}$~\ergl, are likely to produce ultraluminous
 X-ray sources.
The ULX X-1 in M82 is believed to be of this type (Strohmayer \&  Mushotzky  2003; Agrawal \& Misra 2006; 
Dewangan,  Titarchuk, \& Griffiths 2006; Mucciarelli et al. 2006). 
Luminous sources~\#24, 41, 47 and 49 are all associated with either young star clusters or bright starforming complexes. These ULXs are potential IMBH systems, which have been discussed in sections 3.4.4, 3.4.7, 3.4.8 and 3.4.9, respectively.

From the comparison of X-ray luminosities of ULXs detected in NGC~5775 during the \cxo\ and \xmm\ observations, we find that s41 and s47 varied at the most by  less than a factor of two. On the other hand, s24, s28, s40 and s49 displayed dramatic variations. Both these two types of ULXs may form in starforming environments that have sub-solar metalicities. We have already seen that the diffuse gas in the starburst galaxy NGC~5775 is composed of sub-solar materials.
As a result of this the newly-formed massive stars in the disk of NGC~5775 
may leave more massive compact remnants than is typical for 
 solar-metallicity stars. 
The combination of more massive black holes and more massive companions
 can lead to more luminous and longer-lived X-ray emission once mass 
 transfer begins via a stellar wind or through Roche lobe-filling. Wind-fed binaries will
steadily emit X-rays over long time. However, Roche lobe-fill systems will display dramatic
variations both in fluxes and spectra.

\subsection{Relationship between ULXs and  FIR, IR UV luminosities  of interacting galaxies }

From our Chandra survey of ULXs, we found a correlation between the observed number of ULXs per galaxy (N$_{ULX}$) and the far infrared luminosity (L$_{FIR}$).  In addition we have also found that  the observed number of ULXs per galaxy 
and the nearest neighbor distance are anti-correlated (Swartz et al. 2004). This means that the observed number of ULXs  in interacting/merging galaxies strongly depends on L$_{FIR}$. However, Brassington et al. (2005) found a weak correlation between  N$_{ULX}$ and  L$_{FIR}$ of five interacting/merging galaxies. Instead, they found that  L$_{ULX}$, determined following the prescriptions of Colbert et al. (2004), is relatively better correlated with L$_{FIR}$ of the same five galaxies. It may be mentioned that such a correlation is obviously expected because the value of L$_{ULX}$ strongly depends on L$_{FIR}$ (Colbert et al. 2004). Thus, we decided to study the relation between N$_{ULX}$ and L$_{ULX}$ using The Cartwheel (AM 0035-335), M82 (NGC~3034), NGC~3256, Arp~270 (NGC~3395/3396), Arp~299 (NGC~3690/IC~694), Antennae NGC~4038/4039), 
NGC~4485/4490, The Mice (Arp~242, NGC~4676A/B), NGC 5194/5195, NGC~5774/5775, Arp~220 (IRAS 15327+2340), and NGC~7252, which are interacting/merging galaxies with  elevated number of ULXs. To maintain uniformity, we have analyzed  all the archival data from  Chandra and XMM-Newton observations  and detected X-ray sources within the D$_{25}$ isophotes of these galaxies.  When the intrinsic luminosities of the X-ray sources are more than 10$^{39}$~\ergl\ in the 0.5-8.0~keV band, then those objects have been identified as ULX candidates. Next, we computed the number of possible background AGNs within the D$_{25}$ isophotes using the log($N$)--log($S$) relation 
deduced from the \cxo\ Deep Field South survey (Rosati \etal\ 2002).
 Finally, we determined total ULXs per galaxy, which will be considered as N$_{ULX}$, after subtracting the background AGNs.  L$_{ULX}$ was computed using equation 4 of Colbert et al. (2004).  The values of L$_{K}$ were derived using the the K-band apparent magnitude from 2MASS and following the expression given in Seigar (2005). We have used GALEX data or the XMM-Newton/OM data in the UV1 and UV2 bands to determine the values of L$_{UV}$ and that for FIR were derived using the 60 and 100~$\micron$ IRAS flux densities (Devereux \& Eales 1989).  Fig 22 displays the plot of N$_{ULX}$ versus Log (L$_{ULX}$) and fitting gives a slope of 1.03 with a correlation coefficient of 0.34. However, if we exclude  The Cartwheel and Arp~220 from fitting, then the slope changes to 1.38 with a correlation coefficient of 0.83. This fit is shown with the straight line in Fig. 22 and indicates that 
the number of ULXs per galaxy depends on L$_{ULX}$, which is a function of FIR, near-IR and UV luminosities. This is an interesting result and suggests that the value of N$_{ULX}$ depends not only on the SFR but also on the mass of the host galaxies (Colbert et al. 2004). It can also be seen from this figure  that the large number of ULXs detected in The Cartwheel may not be the true ULX population of this galaxy system. At least four to five sources are in excess compared to the expected value. Detection of these four-five sources may be the result of source confusion. On the other hand a large number of ULXs are missing in Arp~220. These missing ULXs may be highly variable, which were below the detection level during the Chandra observations of this galaxy. It is also evident from Fig. 22 that at least two to three ULXs are missing in NGC~~5775. This indicates that the two ULX candidates (s18 and s23) marked as possible halo-objects of NGC~5775, may really belong to this galaxy. Finally, we conclude that the interacting/merging galaxies host comparable number of ULXs based on their mass and starformation rate.

\begin{center}
\includegraphics[angle=-90,width=\columnwidth]{N_ULX_L_ULX.ps}
\figcaption{Plot of  N$_{ULX}$ versus Log (L$_{ULX}$). Solid line shows the fit with slope  1.38 and correlation coefficient of 0.83. This fit is without The Cartwheel and Arp~220.}
\end{center}


\section{Conclusions}
The number of nearby interacting galaxies that exhibit elevated number of ULXs is quite small, so the interacting galaxy pair NGC~5774/5775 is an important system in that regard. These two galaxies are connected through two bridges, which is shown in Fig. 1. The size of the image in Fig. 1 is 6.'6$\times$5.'8 that contains two galaxies, two bridges and adjacent fields in between. We have performed a multiwavelength study of the whole system and the main results are summarized as follows:

1. In total forty nine X-ray sources were detected in the 6.'6$\times$5.'5 field. In this field we expect 3 background sources at or above the detection flux level of a ULX in NGC~5774/5775, based on the log($N$)--log($S$) relation 
deduced from the \cxo\ Deep Field South survey (Rosati \etal\ 2002). However, two and three background AGNs were identified in this field from our optical spectroscopic and photometric studies, respectively.

2. We have detected two sources on the two bridges and these two sources are expected to be HMXRBs. It is likely that interaction-induced starformation was triggered on the bridges and these HMXRBs were formed. Other relatively faint X-ray population  on the bridges  can be detected with deeper X-ray observations. Outside the D$_{25}$ isophote of NGC~5775 but within the 
6.'6$\times$5.'5 field we have detected two ULX candidates, s18 and s23, which appear to be located on the bridges. However, their closeness to the D$_{25}$ isophote of NGC~5775 indicate that they may be in the halo of this edge-on galaxy or at the interface of the bridge and the halo of NGC~5775. In addition, we have detected another bright ($\sim$10$^{41}$~\ergl) ULX, s25, which is associated with a background galaxy SDSS J145355.82+033431.8 at a redshift of 0.0953$\pm$0.0002.

3. Eight sources, excluding one AGN, was detected in an extremely low starforming galaxy, NGC~5774. X-ray population of this galaxy can not be explained using the universal XLF of HMXRBs proposed by Grimm, Gilfanov, \& Sunyaev (2003a). However, it has been demonstrated that the X-ray source population of NGC~5774 consists of the older  and the younger stellar populations, which are directly related to the mass and SFR of NGC~5774 (Colbert \etal\ 2004; Grimm, Gilfanov, \& Sunyaev 2003b).
On the other hand twenty four X-ray sources, excluding one AGN, of the starburst galaxy NGC~5775 can be well described with the universal XLF formalism of  Grimm, Gilfanov, \& Sunyaev (2003a). In fact, the XLF of NGC~5775 is consistent with that of other interacting galaxies. These results indicate that the interacting/merging galaxies   have comparable  numbers of luminous sources (Grimm, Gilfanov, \& Sunyaev 2003a,b; Colbert \etal\ 2004).

4. Presence of AGNs were not detected from the optical spectra of the nuclei of NGC~5774 and NGC~5775. In addition, no nuclear point sources were detected in the Chandra images of these two galaxies suggesting that they  do not host active nuclei.
Diffuse emission was not detected in NGC~5774 but single temperature sub-solar gas is present in NGC~5775. The temperature of this gas ($\sim$0.3 keV) is lower than that to the detected in other interacting galaxies ($\sim$0.5 keV). Un-resolved X-ray sources contribute significantly to the diffuse emission of NGC~5775.
Outflows/winds  are detected in the HST/ACS H$\alpha$ image of NGC~5775, which   suggest that the gas in the  central region of this starburst galaxy is being compressed to produce galactic winds. All these observed properties clearly demonstrate  that NGC~5775 is in the beginning of the evolutionary process  (Taniguchi et al. 2000).

5. Excluding the background AGNs and the two ULX candidates in the halo of NGC~5775, we have detected 10 ULX candidates within the 
D$_{25}$ isophotes of NGC~5774 and NGC~5775. Interestingly, these 10 ULX candidates fall in three groups - (i) s01, s12, s28 and s40 do not have any optical counterparts and are powerlaw X-ray luminous sources with photon indices around 1.8, (ii) s04 and s46 are having steep-powerlaw X-ray spectra and embedded in diffuse H$\alpha$ emission, which are probably ionized nebulae, and (iii) s24, s41, s47 and s49 are hosted in either young star clusters or bright starforming complexes and all are flat-powerlaw X-ray sources. Variable radio emissions from s47 and s49 are detected. ULXs candidates in the first group are believed to be stellar-mass black hole systems. Among these four objects, s40 is the brightest ($\sim$10$^{41}$~\ergl) with a possible 6.2 hr period  and it varied at least by more than a factor of 500. Most likely it is a relativistically beamed ULX and may be similar to Cygnus X-3 (Ghosh et al. 2006). Formation of H$\alpha$ nebulae around s04 and s46 could be due to energetic supernova explosions or to continuous inflation by jets. Variability, high S/N spectra and optical morphology of the whole system will help to determine the nature of these two ULX candidates. Finally, the four ULX candidates of the last group are potential IMBH candidates. Simultaneous radio and X-ray observations of these objects will be extremely useful to determine the mass of the accretors of these ULXs.

6. Number of ULXs in interacting/merging galaxies are correlated with the FIR, K-band and UV luminosities of their host galaxies, suggesting that the formation and evolution of ULXs depend not only on the starformation rate but also on the mass of these galaxies.

\begin{acknowledgements}
Our sincere thanks to the referee's helpful comments and suggestions which improved this paper.
This research has made use of the NASA/IPAC Extragalactic Database (NED) which
 is operated by the Jet Propulsion Laboratory, California Institute of
 Technology, under contract with NASA;
of data products from the Two Micron All Sky Survey, which is a joint project 
 of the University of Massachusetts and the Infrared Processing and Analysis 
 Center, funded by NASA and the NSF;
from the Multimission Archive (MAST) at the STScI operated by AURA under NASA 
 contract NAS5-26555;
and from the Chandra Data Archive, part of the Chandra X-Ray Observatory.
 Science Center (CXC) which is operated for NASA by SAO. The Australia Telescope is 
funded by the Commonwealth of Australia for operation as a National Facility managed
by CSIRO.
Support for this research was provided in part by NASA under 
 Grant NNG04GC86G issued through the Office of Space Science.
We are grateful to Conrado Carretero who made the spectroscopic 
observations at the TNG. The Italian Telescopio Nazionale Galileo (TNG) operated on the island of La Palma by the Fundación Galileo Galilei of the INAF (Istituto Nazionale di Astrofisica) at the Spanish Observatorio del Roque de los Muchachos of the Instituto de Astrofisica de Canarias".  M. L. C. and C. M. G. was supported  by the Program   Ramón y Cajal of the Spanish science ministery.

\end{acknowledgements}


Neff, Ulvestad \& Campion 2003; Kaaret et al. 2003; Kording, Colbert \& Falcke 2005; Miller, Mushotzky \& Neff 2005; Soria et al. 2006; Lang et al. 2007






\clearpage

\begin{center}
\scriptsize{
\begin{tabular}{ccccccccrcc}
\multicolumn{11}{c}{{\sc Table 1}} \\
\multicolumn{11}{c}{X-ray sources in the field of the interacting--pair of galaxies, NGC5774/5775} \\
\hline \hline
\multicolumn{1}{c}{Source}& \multicolumn{1}{c}{R.A.} & \multicolumn{1}{c}{Dec.} & \multicolumn{1}{c}{Counts}& \multicolumn{1}{c}{N$_{H}$}&\multicolumn{1}{c}{$\Gamma$} &\multicolumn{1}{c}{$\chi^{2}$/dof} &\multicolumn{1}{c}{L$_{X}$}&\multicolumn{1}{c}{F$_{X}$/F$_{O}$}&\multicolumn{2}{c}{Variability}\\
(\#)&\multicolumn{1}{c}{(2000)} &\multicolumn{1}{c}{(2000)} &\multicolumn{1}{c}{(58.2~ks)} &\multicolumn{1}{c}{(10$^{21}$~cm$^{-2}$)}&&&\multicolumn{1}{c}{(10$^{39}$~ergs/s)}&&\multicolumn{1}{c}{P$_{KS}$}&\multicolumn{1}{c}{$\chi^{2}$(59 dof)}\\
\hline
1 &14 53 39.948 &3 34 19.026 &216 &3.19$_{-1.25}^{+1.99}$&1.86$_{-0.28}^{+0.31}$&14.3/15&2.83$\pm$0.19&$>$16.81&0.000&  85.7\\
2 &14 53 42.222 &3 31 42.792 &32 &0.347&1.8&---&0.33$\pm$0.05&0.14&0.865&40.8\\
3 &14 53 42.762 &3 35 3.222 &15 &0.347&1.8&---&0.15$\pm$0.04&--&0.378&19.7\\
4 &14 53 43.776 &3 34 27.39 &69 &4.49$_{-4.49}^{+12.5}$&1.64$_{-1.53}^{+2.75}$&0.47/2&1.32$\pm$0.23&--&0.001&37.3\\
5 &14 53 44.694 &3 33 30.834 &846 &0.54$_{-0.51}^{+0.43}$&2.08$_{-0.17}^{+0.24}$ &58.2/59&6.57$\pm$0.34&0.31&0.731&54.7\\
6 &14 53 44.856 &3 32 57.48 &41 &0.347& 1.71$_{-0.76}^{+1.00}$&2.5/3&0.39$\pm$0.06&0.09&0.758&44.8\\
7 &14 53 45.18 &3 33 48.648 &30 &0.347&1.8&---&0.29$\pm$0.05&$>$1.90&0.986&35.3\\
8 &14 53 46.254 &3 33 50.658 &16 &0.347&1.8&---&0.17$\pm$0.04&$>$0.89&0.191&37.4\\
9 &14 53 46.5 &3 34 59.352 &16 &0.347&1.8&---&0.16$\pm$0.02&0.06&0.473&22.7\\
10 &14 53 46.644 &3 32 57.564 &23 &0.347&1.8&---&0.22$\pm$0.03&$>$1.42&0.055&44.1\\
11 &14 53 47.88 &3 34 5.07 &19 &0.347&1.8&---&0.18$\pm$0.02&0.04&0.928&31.6\\
12 &14 53 48.282 &3 34 2.448 &117 &1.31$_{-1.31}^{+2.62}$&1.40$_{-0.48}^{+0.75}$&8.79/9&1.77$\pm$0.22&$>$10.52&0.123&71.3\\
13 &14 53 48.63 &3 31 58.878 &39 &0.347&1.8&---&0.40$\pm$0.03&0.41&0.877&40.4\\
14 &14 53 51.9 &3 35 24.216 &38 &0.347&1.8&---&0.37$\pm$0.02&$>$2.37&0.443&50.6\\
15 &14 53 52.308 &3 33 17.982 &27 &0.347&1.8&---&0.27$\pm$0.03&0.14&0.047&48.0\\
16 &14 53 52.506 &3 34 5.982 &24 &0.347&1.8&---&0.24$\pm$0.06&$>$1.42&0.089&39.0\\
17 &14 53 52.65 &3 31 40.554 &44 &0.83$_{-0.83}^{+0.62}$&1.58$_{-0.73}^{+1.67}$&1.12/2&0.53$\pm$0.17&$>$3.29&0.771&37.4\\
18 &14 53 53.13 &3 32 55.578 &75 &9.25$_{-9.25}^{+17.0}$&1.30$_{-0.79}^{+1.32}$&5.75/5&2.06$\pm$0.53&$>$12.24&0.259&44.1\\
19 &14 53 53.706 &3 36 23.994 &94 &1.97$_{-1.47}^{+2.17}$&1.89$_{-0.61}^{+0.62}$&3.79/6&1.08$\pm$0.13&$>$6.41&0.051&40.1\\
20 &14 53 54.324 &3 33 49.884 &40 &0.347&1.8&---&0.41$\pm$0.05&0.01&0.444&55.0\\
21 &14 53 54.75 &3 33 40.656 &55 &1.30$_{-1.30}^{+32.98}$&1.88$_{-0.97}^{+6.46}$&9.45/5&0.58$\pm$0.08&--&0.511&50.6\\
22 &14 53 55.146 &3 32 45.834 &23 &0.347&1.8&---&0.24$\pm$0.02&--&0.447&37.5\\
23 &14 53 55.248 &3 32 29.178 &182 &3.04$_{-1.84}^{+1.41}$&2.49$_{-0.43}^{+0.56}$&4.31/13&1.96$\pm$0.27&1.49&0.837&66.9\\
24 &14 53 55.758 &3 33 28.068 &252 &12.27$_{-5.37}^{+8.21}$&1.28$_{-0.32}^{+0.69}$&13.62/20&7.78$\pm$0.32&--&0.378&49.8\\
25 &14 53 55.896 &3 34 30.258 &34 &0.347&1.8&---&0.35$\pm$0.04&$>$1.60&0.854&40.9\\
26 &14 53 55.92 &3 34 0.72 &355 &0.94$_{-0.74}^{+0.73}$&1.29$_{-0.15}^{+0.21}$&21.22/27&4.63$\pm$0.61&0.21&0.088&83.6\\
27 &14 53 56.256 &3 33 3.132 &30 &0.347&1.8&---&0.29$\pm$0.06&--&0.478&39.3\\
28 &14 53 56.778 &3 33 8.55 &124 &1.57$_{-1.57}^{+4.21}$&1.83$_{-0.64}^{+0.61}$&15.5/8&1.34$\pm$0.17&--&0.308&39.7\\
29 &14 53 56.796 &3 31 29.346 &17 &0.347&1.8&---&0.17$\pm$0.02&$>$7.96&0.623&46.9\\
30 &14 53 56.844 &3 32 40.878 &15 &0.347&1.8&---&0.14$\pm$0.05&--&0.492&46.9\\
31 &14 53 56.892 &3 32 57.582 &51 &2.14$_{-2.14}^{+7.16}$&2.15$_{-0.98}^{+1.77}$&19.26/4&0.66$\pm$0.17&--&0.812&50.3\\
32 &14 53 57.156 &3 32 51.648 &27 &0.347&1.8&---&0.27$\pm$0.08&--&0.146&39.1\\
33 &14 53 57.348 &3 32 43.212 &23 &0.347&1.8&---&0.22$\pm$0.04&--&0.741&32.3\\
34 &14 53 57.42 &3 32 26.394 &15 &0.347&1.8&---&0.33$\pm$0.07&--&0.719&45.9\\
35 &14 53 57.618 &3 32 41.85 &26 &0.347&1.8&---&0.27$\pm$0.07&--&0.488&48.3\\
36 &14 53 57.648 &3 32 36.51 &15 &0.347&1.8&---&0.14$\pm$0.04&--&0.061&38.8\\
37 &14 53 57.894 &3 32 38.502 &19 &0.347&1.8&---&0.18$\pm$0.04&--&0.475&27.7\\
38 &14 53 57.96 &3 32 17.514 &11 &0.347&1.8&---&0.10$\pm$0.05&--&0.343&42.1\\
39 &14 53 58.668 &3 35 6.342 &14 &0.347&1.8&---&0.13$\pm$0.02&0.10&0.426&27.7\\
40 &14 53 58.896 &3 32 16.788 &1354 &29.6$_{-3.92}^{+5.00}$&1.99$_{-0.24}^{+0.29}$&112.9/110&75.84$\pm$7.23&--&0.495&43.5\\
41 &14 53 59.448 &3 31 57.24 &307 &4.59$_{-1.43}^{+2.41}$&1.48$_{-0.24}^{+0.32}$&20.28/28&6.06$\pm$0.25&--&0.723&56.1\\
42 &14 53 59.478 &3 31 47.79 &26 &0.347&1.8&---&0.27$\pm$0.05&--&0.642&45.5\\
43 &14 53 59.754 &3 31 40.284 &37 &0.347&1.8&---&0.36$\pm$0.03&--&0.383&44.1\\
44 &14 54 0.144 &3 31 31.05 &15 &0.347&1.8&---&0.14$\pm$0.04&--&0.325&31.6\\
45 &14 54 0.81 &3 31 30.96 &11 &0.347&1.8&---&0.10$\pm$0.03&--&0.022&34.4\\
46 &14 54 0.96 &3 31 29.43 &92 &17.5$_{-10.8}^{+13.4}$&2.30$_{-0.70}^{+0.97}$&10.28/7&3.22$\pm$0.60&--&0.011&42.3\\
47 &14 54 0.972 &3 31 33.072 &125 &14.3$_{-6.21}^{+12.3}$&1.19$_{-0.89}^{+0.45}$&6.86/9&4.42$\pm$0.92&--&0.221&55.8\\
48 &14 54 2.964 &3 32 10.056 &21 &0.347&1.8&---&0.21$\pm$0.04&$>$1.31&0.842&45.9\\
49$^{a}$ &14 53 59.87 & 3 32 18.0 &&0.6$_{-0.60}^{+0.60}$&1.49$_{-0.44}^{+0.23}$&&5.04$\pm$0.56&\\
\hline
\multicolumn{11}{l}{$^a$Detected only during the XMM-Newton observations.}\\ 
\end{tabular}
} 
\end{center}

\clearpage

\begin{center}
\scriptsize{
\begin{tabular}{ccccc}
\multicolumn{5}{c}{{\sc Table 2}} \\
\multicolumn{5}{c}{Spectral parameters of XMM-Newton sources in NGC~5774/5775} \\
\hline \hline
\multicolumn{1}{c}{Source}&  \multicolumn{1}{c}{$\Gamma$} &\multicolumn{1}{c}{N$_{H}$} &\multicolumn{1}{c}{$\chi^{2}$/dof}& \multicolumn{1}{c}{L$_{X}$}\\
\#&&\multicolumn{1}{c}{(10$^{21}$~cm$^{-2}$)}&&\multicolumn{1}{c}{(10$^{39}$~ergs/s)}\\
\hline
5$^a$&1.49$^{+0.23}_{-0.44}$& 0.56$^{+0.60}_{-0.56}$&20.4/25&5.1$\pm$0.6\\
19&1.78$^{+0.26}_{-0.24}$& 3.1$^{+0.43}_{-0.66}$&9.1/12&1.8$\pm$0.4\\
23&2.15$^{+0.39}_{-0.40}$& 2.9$^{+0.49}_{-0.29}$&19/21&2.4$\pm$0.5\\
26$^a$&1.98$^{+0.12}_{-0.28}$& 3.1$^{+1.3}_{-0.8}$&38.9/31&6.3$\pm$0.94\\
41&1.40$^{+0.17}_{-0.14}$&2.8$^{+1.1}_{-0.7}$&52.6/46&12.65$\pm$1.45\\
47&0.45$^{+0.25}_{-0.29}$&0.347&8.5/12&7.53$\pm$0.70\\
49&1.49$^{+0.23}_{-0.44}$&0.6$^{+0.6}_{-0.6}$&20.4/22&5.04$\pm$0.56 \\
\hline
\multicolumn{5}{l}{$^a$Estimated ROSAT/HRI luminosities for sources 5 and 26 are }\\ 
\multicolumn{5}{l}{(6.0$\pm$1.0) and (5.0$\pm$1.0) $\times$ 10$^{39}$ ~ergs~s$^{-1}$, respectively.}\\
\end{tabular}
} 
\end{center}

\clearpage

\begin{center}
\scriptsize{
\begin{tabular}{ccccccc}
\multicolumn{7}{c}{{\sc Table 3}} \\
\multicolumn{7}{c}{Photometric magnitudes of optical counterparts of Chandra sources} \\ 
\multicolumn{7}{c}{detected in the field of NGC~5774/5775} \\
\hline \hline
\multicolumn{1}{c}{Source}& \multicolumn{5}{c}{SDSS photometric magnitudes}& \multicolumn{1}{c}{Remarks}\\ 
\#&\multicolumn{1}{c}{u}& \multicolumn{1}{c}{g}&\multicolumn{1}{c}{r}&\multicolumn{1}{c}{i}&\multicolumn{1}{c}{z}&\\
\hline
2  &24.77& 22.46 &22.01 &22.03& 21.80&background object$?$\\ 
5  &20.90& 20.25 &19.47 &19.09& 18.87 &QSO at z=1.365\\
6  &22.71& 21.63 &21.30 &21.42 &21.19&background AGN$?$\\ 
9  &23.46& 22.64 &21.73& 21.19& 21.26&background AGN$?$\\
11 &23.69 &21.79& 21.04 &20.74 &20.44&globular cluster$?$\\ 
13 &23.03 & 22.92  &23.21  &23.39  &22.12&unknown object\\ 
15 &21.92& 20.92& 20.60 &20.73 &20.19&background AGN$?$\\ 
17 &22.62& 23.53& 23.73 &22.79 &22.47&blue compact dwarf galaxy\\ 
23 &24.83 &22.80 &22.77 &22.93 &21.49&globular cluster$?$\\ 
25 &22.98 &23.87 &21.80 &24.24 &20.78 &ULX in a Galaxy at z=0.0953\\
26 &20.88& 20.21 &19.43 &19.07 &18.85 &QSO at z=0.661\\
39 &23.47 &24.20 &21.83 &20.96 &20.15&peculiar object\\ 
\hline
\end{tabular}
} 
\end{center}

\end{document}